\documentclass[twocolumn,prb,citeautoscript,floatfix,final,aps]{revtex4}
\usepackage{graphicx}
\usepackage{amsmath,amssymb,amsfonts}
\usepackage{gensymb}
\usepackage{bm}

\usepackage[T1]{fontenc}
\usepackage{textcomp}
\usepackage{mathptmx}
\usepackage[scaled=0.92]{helvet}
\usepackage{courier}
\newcommand{\Leff}{\ensuremath{L_{\textit{eff}}}}
\newcommand{\heff}{\ensuremath{h_{\textit{eff}}}}
\newcommand{\hefftilde}{{\ensuremath{\widetilde{h}_{\textit{eff}\,}}(\gamma)}}
\newcommand{\Ltip}{\ensuremath{L_{\textit{tip}}}}
\newcommand{\lside}{\ensuremath{l_{\textit{side}}}}
\newcommand{\lproj}{\ensuremath{l_{\textit{proj}}}}
\newcommand{\Sf}{\ensuremath{{(sf)}}}
\newcommand{\Ss}{\ensuremath{{(ss)}}}
\newcommand{\Max}{\ensuremath{{\textit{max}}}}
\newcommand{\Min}{\ensuremath{{\textit{min}}}}
\renewcommand*{\vec}[1]{\mathbf{#1}}
\newcounter{summary}
\newcommand{\sumnum}{\stepcounter{summary}\par\thesubsection.\arabic{summary}. }
\newcommand{\subsumnum}{\stepcounter{summary}\par\thesubsection.\thesubsubsection.\arabic{summary}. }
\begin{document}
%
\title{Critical adsorption and critical Casimir forces for geometrically structured confinements}
%
\author{M.~Tr{\"o}ndle}
\author{L.~Harnau}
\author{S.~Dietrich}
\affiliation{
	Max-Planck-Institut f\"ur Metallforschung,  
	Heisenbergstr.\ 3, D-70569 Stuttgart, Germany, 
	\\
	and Institut f\"ur Theoretische und Angewandte Physik, 
	Universit\"at Stuttgart, 
	Pfaffenwaldring 57, 
	D-70569 Stuttgart, Germany
}
\date{August 14, 2008}
%
%
\begin{abstract}
	We study the behavior of fluids, confined by geometrically structured substrates,
	upon approaching a critical point at ${T=T_c}$ in their bulk phase diagram.
	As generic substrate structures
	periodic arrays of wedges and ridges are considered.
	Based on general renormalization group arguments we calculate, within mean field approximation, the
	universal scaling functions for
	order parameter profiles of a fluid close to a single structured substrate and discuss
	the decay of its spatial variation into the bulk.
	We compare the excess adsorption at corrugated substrates with the one at planar
	walls.
	The confinement of a critical fluid by two walls generates effective critical Casimir forces
	between them.
	We calculate corresponding universal scaling functions for the normal critical Casimir force
	between a flat and a geometrically structured substrate
	as well as the lateral critical Casimir force between two
	identically patterned substrates.
\end{abstract}
\maketitle
%
%
\section{Introduction}
%
The confinement of a fluid which is close to its critical point at ${T=T_c}$ induces remarkable deviations 
from its bulk behavior.
Typically, the boundary of a system affects the local structural properties of condensed matter within a
layer of thickness of the bulk correlation length,
which diverges upon approaching criticality as
$\xi_\pm(t\to0)=\xi_0^{\pm}|t|^{-\nu}$, where $t=(T-T_c)/T_c$,
$\nu$ is a standard bulk critical exponent, 
and
$\xi_0^\pm$ are nonuniversal amplitudes above ($+$) or below ($-$) $T_c$.
For binary liquid mixtures, the disordered phase at ${T>T_c}$ is the one in which its two components are mixed, whereas
in the ordered phase at ${T<T_c}$ there is phase separation in two phases being rich in one or the other component.
(At lower critical points the disordered phase occurs for ${T<T_c}$ and the ordered ones at ${T>T_c}$.)
Accordingly, near the critical demixing point $T_c$
the scalar order parameter $\phi$ is a suitably defined concentration difference
of the two species,
belonging to the Ising universality class.
The generic preference of a surface for either one of the two species forming the binary fluid 
amounts to the presence of an effective surface field which
affects the order parameter profile close to the wall such that it is nonzero even in the 
disordered (mixed) phase for ${T>T_c}$.
Upon approaching $T_c$, these so-called critical adsorption profiles,
describing the concentration enhancement near the surface, become long ranged due to the divergence
of the correlation length \cite{FISHER78,BINDER83,DIEHL86}.
In the semi-infinite geometry
the transitions from the phase in which only the region near a single surface is ordered 
to the one in which also the bulk is ordered are known as the so-called
extraordinary or normal surface transitions \cite{BRAY77,BURKHARDT94}.
In the slab geometry the confinement of critical fluctuations of the order parameter gives rise to critical Casimir forces
attracting or repelling the confining walls, which can be of significant strength \cite{FISHER78,KRECH91,KRECH92a,KRECH92b,KRECH94,KRECH97}.
This is the thermodynamic analogue of the quantum-electrodynamic Casimir effect originating from the confinement of vacuum
fluctuations \cite{CASIMIR48,CAPASSO07}.
Critical adsorption has been investigated experimentally for flat substrates (see, e.g., 
Refs.~~\onlinecite{FLOETER95,LAW2001,LIU89,Brown:2007,Brown:2007a} and references therein).
Also, there is strong experimental evidence for the occurrence of critical Casimir forces 
\cite{CHAN02,Fukuto:2005,Ganshin:2006,Rafai:2007}, and recently a direct measurement of the 
critical Casimir force between a wall and a colloid has been reported \cite{Hertlein:2008}.
So far the theoretical investigations of critical adsorption and the critical Casimir effect have been focused on topologically
flat surfaces (see, e.g., Refs.~~\onlinecite{BINDER83,DIEHL86,LAW2001,KRECH97,Vasilyev:2007,Hucht:2007}) or curved surfaces and colloids
(see, e.g., Refs.~~\onlinecite{Burkhardt:1995,Eisenriegler:1995,Hanke:1998,Schlesener:2003,Eisenriegler:2004,SLAVA07A}).
\par
Nowadays, experimental techniques are available which allow one to endow a solid surface
with a precisely defined {geometrical} or chemical structure in the
micro- and nanometer range (see, e.g., Refs.~~\onlinecite{Xia:1998,Wang:1998,Tolfree:1998,Nie:2008}).
Chemically and geometrically structured substrates are of major importance for 
the construction of a 'lab on a chip' \cite{Thorsen:2002}.
Critical adsorption and critical Casimir forces for chemically 
structured but topologically flat confinements have been studied theoretically \cite{SPRENGER05, SPRENGER06B}.
Typical man-made geometrical structures are periodic in one direction and consist of grooves 
with wedge-like shapes.
The effects of such roughness and of the structuring of substrates, for example, on wetting phenomena 
(see, e.g., Refs.~~\onlinecite{Cassie:1944,MIKO06,MIKO07} and references therein), liquid crystal systems 
\cite{Harnau:2004,Harnau:2006,HARNAULC,Karimi:2006}, or depletion interactions \cite{Zhao:2007} have been found to be crucial.
\par
In view of this context, as a paradigmatic first step critical adsorption at a single wedge and a single
ridge has been studied at the critical point \cite{HANKE99} and off-criticality \cite{PALAGYI04}.
It was found that, in agreement with earlier work \cite{CARDY83}, the angle characterizing the wedge is a 
parameter which influences the critical behavior significantly.
\par
On the other hand, the effect of geometric structures of surfaces on the quantum electrodynamic Casimir 
effect has been analyzed, too \cite{Emig:2001,EMIG03,Zwol:2008,Chan:2008,Lambrecht:2008}.
Periodic structures give rise to lateral quantum electrodynamic Casimir forces \cite{Chen:2002,EMIG03,
EMIG05,Rodrigues:2006,Dalvit:2008}. 
Recent studies have been focused on quantum electrodynamic systems in which lateral Casimir forces due to 
topological structures lead to a controlled motion \cite{MIRI07,EMIG07,Miri:2008,Rodrigues:2008}.
\par
The present work is supposed to extend these investigations into various directions.
As a paradigmatic model for a geometrically structured substrate, we consider a periodic array
of wedges and ridges.
One expects an interesting interplay between the externally endowed structures and the critical
behavior of the system if the size scales of the substrate corrugation are comparable with
the correlation length. 
Actually, this range is experimentally accessible because the correlation length can 
reach values up to several hundreds of nanometers (e.g., for a mixture of water and lutidine 
one finds $\xi_0^+\simeq 0.2$\,nm so that for $t=10^{-4}$, corresponding to  $T-T_c\simeq0.03K$ in that system,
one has $\xi\simeq 70$\,nm \cite{Gulari:1972,Hertlein:2008}).
First, we describe the critical adsorption behavior of the order parameter profile for the whole temperature 
range around $T_c$ in terms of universal scaling functions based on general renormalization group arguments.
We calculate these corresponding universal scaling functions up to lowest order, i.e., within mean field theory,
and subsequently we discuss the experimentally relevant excess adsorption.
Second, we study the critical Casimir effect for geometrically structured confinements and focus 
on the universal features of the normal and lateral forces emerging in binary
liquid mixtures close to criticality.
We calculate universal scaling functions for the critical Casimir forces within mean field approximation
for identical chemical boundary conditions on both walls ($(+,+)$ configuration).
\par
The paper is organized as follows. 
Section~\ref{sec:criticaladsorption} is devoted to the study of critical adsorption on a single
geometrically structured substrate.
General scaling properties of the order parameter profile are given in Subsec.~\ref{sec:opmgeneral},
followed by a detailed investigation of mean field results for the corresponding universal scaling 
functions in Subsec.~\ref{sec:opm}.
In Subsec.~\ref{sec:excess} we study the excess adsorption at corrugated surfaces.
Section~\ref{sec:casimir} discusses critical Casimir forces between geometrically structured walls mediated
by the enclosed critical fluids.
In particular, Subsec.~\ref{sec:normal} addresses the  normal force between a flat and a geometrically 
structured substrate, and
in Subsec.~\ref{sec:lateral} we discuss the lateral and the normal critical Casimir force between 
two identically structured substrates.
Section~\ref{sec:summary} summarizes our findings.
%

%
%
\section{Critical adsorption on geometrically structured substrates \label{sec:criticaladsorption}}
%
\subsection{Model for geometrically structured substrates\label{sec:periodicarray}}
\begin{figure}

	\includegraphics{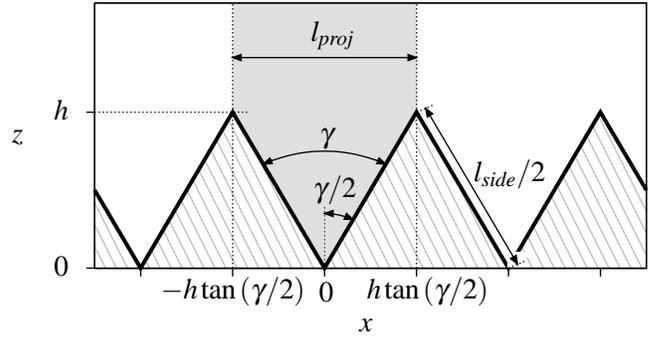}
	\caption{%
		Side view on a periodic array of wedges and ridges modeling a geometrically structured substrate.
		The structure is periodic along the $x$ direction and the shaded region indicates one unit cell.
		The direction globally perpendicular to the substrate is denoted as $z$.
		The structure is infinitely extended within the $x$-$\vec{y}$ plane, where $\vec{y}$ are
		the directions in which the structure is translationally invariant.
		}
	\label{fig:definitions}
\end{figure}

We model a laterally corrugated substrate confining a fluid by a periodic array of symmetric wedges and
ridges as shown in Fig.~\ref{fig:definitions}.
%
%
The direction along which the structure varies periodically is denoted
as $x$, the
direction indicating the distance from the substrate is called $z$, and the remaining $(d-2)$
directions in which the system is translationally invariant are denoted as $\vec{y}$.
Since for this system the observables do not depend on $\vec{y}$, we are left with
an effective two-dimensional problem.
The geometric structure of the confinement is characterized (Fig.~\ref{fig:definitions}) 
by the corrugation height $h$ and the wedge angle $\gamma$.
The corrugation wavelength is given by $\lproj=2h\tan(\gamma/2)$ and the actual
substrate surface area is proportional to $\lside=2h/cos(\gamma/2)$.
The projected area $A$ of the substrates in the $(d-1)$ directions $\{x,\vec{y}\}$ is 
macroscopically large
(the notation $A\to\infty$ will be suppressed in the following).
\par
One has to stress that all universal, system-independent quantities or functions discussed below reliably 
describe actual physical quantities only for distances from the walls larger than typical molecular 
length scales of the system, i.e., the length scales of the constituent particles of the fluid and of the confining substrates.
This means that all relevant length scales must be much larger than a typical molecular size.
For the correlation length $\xi$ this requirement is fulfilled near criticality.
For the geometric structure of the substrate this requirement implies that both $h$ and also
$\lproj$ must be much larger than a characteristic, system-dependent molecular length scale.
One should keep in mind that the requirement that $\lproj$ is larger than molecular sizes limits the 
range of validity of the universal properties for critical adsorption for small wedge angles $\gamma\to0$.
%
\subsection{General scaling properties for the order parameter profile \label{sec:opmgeneral}} 
%
In this subsection critical adsorption on a {single} geometrically structured substrate is studied.
The order parameter $\phi$ is a function of the spatial coordinates $x$ and $z$, the reduced temperature $t$,
as well as of the corrugation parameters: $\phi=\phi(x,z,h,\gamma,t)$.
Close to the bulk critical temperature $\phi$ takes the scaling form \cite{LIU89,FLOETER95,PALAGYI04,Diehl:1993}
\begin{equation}
	\label{eq:scaling}
	\phi(x,z,h,\gamma,t)=a|t|^\beta P_\pm(x_\pm,z_\pm,h_\pm,\gamma),
\end{equation}
where $x_\pm=x/\xi_\pm$ is the direction of periodicity, $z_\pm=z/\xi_\pm$
is the \emph{global} normal distance from the substrate, and $h_\pm=h/\xi_\pm$  is the
corrugation height. All lengths are scaled by the correlation length.
The scaling functions $P_\pm$ are universal once the nonuniversal amplitudes $a$ and $\xi_0^+$
are fixed as the amplitude of the bulk order parameter and of the true correlation length corresponding to an
exponential decay of the pair correlation function.
For distances from the substrate which are small compared to $\xi_\pm$, or for $T\to T_c$, 
the scaling functions and the order parameter exhibit power law singularities \cite{HANKE99,PALAGYI04}:
\begin{equation}
	\label{eq:scaling-zero}
	\phi(x,z,h,\gamma,t=0)=a\,\widetilde{C}_\pm(x/h,z/h,\gamma)\left[\frac{z}{\xi_0^\pm}\right]^{-\beta/\nu},
\end{equation}
where $\widetilde{C}_\pm$ are dimensionless, universal amplitude functions depending only on distances 
scaled by the corrugation height $h$.
The usage of the subscripts '$\pm$' for the scaling function in Eq.~\eqref{eq:scaling-zero} refers to the fact that they are derived
from the scaling functions $P_\pm$ in the limit $t\to0$ from above or below $T_c$.
Equation~\eqref{eq:scaling-zero} implies the relation $\widetilde{C}_+(x/h,z/h,\gamma)\big/\widetilde{C}_-(x/h,z/h,\gamma)=
(\xi_0^+/\xi_0^-)^{-\beta/\nu}$ between the scaling functions $\widetilde{C}_\pm$ and the universal amplitude ratio of the correlation
lengths above and below $T_c$ \cite{HANKE99,tarko:1973,tarko:1975}.
\subsubsection{Distant behavior}
For distances from the substrate much larger than the correlation length the scaling functions for $T\neq T_c$ decay as
\begin{equation}
	\label{eq:scaling-decay}
	P_+(x_+,z_+\to\infty,h_+,\gamma)	\;\propto\;	\exp\{-z_+\}
\end{equation}
and
\begin{equation}
	\label{eq:scaling-decay-2}
	P_-(x_-,z_-\to\infty,h_-,\gamma)-1	\;\propto\;	\exp\{-z_-\}.
\end{equation}
Generally, for $z/h\to\infty$ it is not possible to detect the corrugation of the confining wall, and
the system resembles a semi-infinite system with an {effective planar} wall.
The leading correction due to the corrugation is that, viewed from a distance, the wall does not
appear as a planar wall located at $z=0$ but located at an {effective} height $z=\heff\le h$ with
\begin{equation}
	\label{eq:def-hefftilde}
	\heff=h\, \hefftilde,
\end{equation}
so that
\begin{equation}
	\label{eq:heff-decay}
	\phi(x,z,h,\gamma,t=0)=a\,C_\pm(x/h,z/h,\gamma)%
	\left[\frac{ z-\heff }{\xi_0^\pm}\right]^{-\beta/\nu},
\end{equation}
where 
\begin{equation}
	\label{eq:definition-function-heff}
	C_\pm(x/h,\,z/h,\,\gamma)=\widetilde{C}_\pm(x/h,z/h,\gamma)\left(1-\frac{h}{z}\;
	\hefftilde\right)^{\beta/\nu}.
\end{equation}
The effective height is implicitly defined by the request that far from the wall
\begin{equation}
	\label{eq:definition-heff}
	C_\pm\left(\frac{x}{h},\frac{z}{h}\to\infty,\gamma\right)=c_\pm,
\end{equation}
where $c_\pm$ are the universal surface amplitudes for the corresponding semi-infinite planar wall system,
defined via \cite{FLOETER95} $\phi^{\infty/2}(z,t=0)=a\, c_\pm\,\left(z/\xi_0^\pm,\right)^{-\beta/\nu}$, where
the superscript $\infty/2$ refers to the semi-infinite system.
Off criticality the exponential decay of the scaling functions is proportional to 
$\exp\{-(z-\heff)/\xi_\pm\}$ so that even in the limit $z\to\infty$ the amplitude of the exponential decay
$\propto \exp(-z/\xi_\pm)$ still carries the information about the corrugation via the prefactor 
$\exp(h\;\hefftilde/\xi_\pm)$.
At $T_c$ the leading decay behavior is $a\,c_\pm\,(z/\xi_0^\pm)^{-\beta/\nu}$, which is independent of the corrugation,
which for $z\to\infty$ leaves its trace only through subdominant algebraic terms.
%
\subsection{Order parameter profiles within mean field theory\label{sec:opm}}
%
Within a field theoretical renormalization group approach the general standard fixed point Hamiltonian 
for critical phenomena in confinements is given by
\cite{BINDER72,BINDER83,DIEHL86}
\begin{equation}
	\label{eq:hamiltonian}
	\mathcal{H}[\phi]=\mathcal{H}_b[\phi]+\mathcal{H}_s[\phi]+\mathcal{H}_e[\phi],
\end{equation}
where the dimensionless bulk Hamiltonian reads
\begin{equation}
	\label{eq:bulkhamiltonian}
	\mathcal{H}_b[\phi]=\int_V\,d^d\textnormal{r}\,\left\{
			 \frac{1}{2}\left(\nabla\phi\right)^2
			+\frac{\tau}{2}\phi^2
			+\frac{u}{4!}\phi^4
		\right\}.
\end{equation}
Equation \eqref{eq:bulkhamiltonian} holds for zero bulk fields as assumed in the following.
The integration in Eq.~\eqref{eq:bulkhamiltonian} runs over the volume $V$ which is accessible
to the fluid, and $\vec{r}=\{x,z,\vec{y}\}$ is the spatial position vector.
In Eq.~\eqref{eq:hamiltonian} the surface Hamiltonian $\mathcal{H}_s$ and the edge Hamiltonian $\mathcal{H}_e$
incorporate the surface (edge) enhancement of the coupling energy at the boundary and
the surface (edge) fields which act on particles close to the surface (edge) only.
At the critical adsorption fixed point \cite{BURKHARDT94}
the surface and edge contributions turn into boundary conditions corresponding to infinite surface and edge fields
describing the strong adsorption limit so that 
$\phi\big|_{\text{surface}}=+\infty$ at all surfaces and edges of
the system.
Since our study deals with identical chemical boundary conditions at all surface and edges,
it is sufficient to consider only the '$+$' case corresponding to $\phi=+\infty$.
In Eq.~\eqref{eq:bulkhamiltonian} the coefficient $\tau$ is proportional to the reduced temperature
$t$, and the coupling constant $u$ is positive.
\par
Quantitative results can be obtained within the framework of mean field theory (MFT),
which corresponds to the zeroth order contribution in a systematic $\varepsilon=4-d$ expansion 
within the field theoretical renormalization group approach.
Within the MFT one neglects fluctuations of the order parameter, and only 
the configuration $m=\sqrt{(u/3!)}\langle\phi\rangle$ with the largest statistical weight 
$\exp(-\mathcal{H}[\phi])$ is considered.
This configuration fulfills the Euler-Lagrange equation
\begin{equation}
	\label{eq:diffeq}
	\nabla^2 m = \tau m + m^3,
\end{equation}
which is obtained via functional minimization of Eq.~\eqref{eq:bulkhamiltonian}.
The universal quantities and scaling functions calculated via MFT are exact
for dimensions larger than the upper critical dimension $d_{uc}=4$.
The mean field values of the critical exponents $\beta$ and $\nu$ are $\beta=\nu=\frac{1}{2}$, and
for the coefficient $\tau$ in Eq.~\eqref{eq:bulkhamiltonian} one has within MFT
$\tau=t(\xi_0^+)^{-2}$ for $t>0$ and $\tau=\frac{1}{2}t(\xi_0^-)^{-2}$ for $t<0$ 
with the universal ratio $(\xi_0^+/\xi_0^-)^2=2$ \cite{tarko:1973,tarko:1975}.
Within MFT and $d=4$ the coupling constant $u$ in Eq.~\eqref{eq:bulkhamiltonian} can be related to 
the nonuniversal constant $a$ in Eq.~\eqref{eq:scaling} as $a=\sqrt{(3!/u)}\big/\xi_0^+$.
For ${T=T_c}$ the dimensionless value  $m\times h$ depends on $x/h$, $z/h$, and $\gamma$ only 
and is independent of the nonuniversal quantities $a$ and $\xi_0^+$ (see Eq.~\eqref{eq:scaling-zero}). 
\par
The order parameter profiles presented in the following have been calculated numerically.
In order to obtain a boundary condition for the numerical calculation we use the short distance expansion 
of the solution of Eq.~\eqref{eq:diffeq} for the semi-infinite system with the boundary condition $m=+\infty$ at the wall 
\begin{equation}
	\label{eq:shortdistance}
	m=\frac{\sqrt{2}}{R}-\frac{\tau}{3\sqrt{2}}R+\mathcal{O}(R^3),
\end{equation}
where $R$ is the minimal distance to the wall.
Eq.~\eqref{eq:shortdistance} holds only for distances $R\ll \xi,h,\lproj$.
We approximate edges by small inscribed circles and use Eq.~\eqref{eq:shortdistance} 
for the numerical boundary condition close to edges with the distance $R$ measuring the minimal distance 
to the edge.
We find the universal scaling functions for the order parameter profiles upon functional minimization
by using a 
conjugate gradient algorithm within a finite element method.
The grids and the numerical boundaries together with the minimization conditions have been chosen such that 
for the order parameter profile the relative error is less than $0.1\%$.
\subsubsection{Comparison of order parameter profiles}
\begin{figure}

	\includegraphics{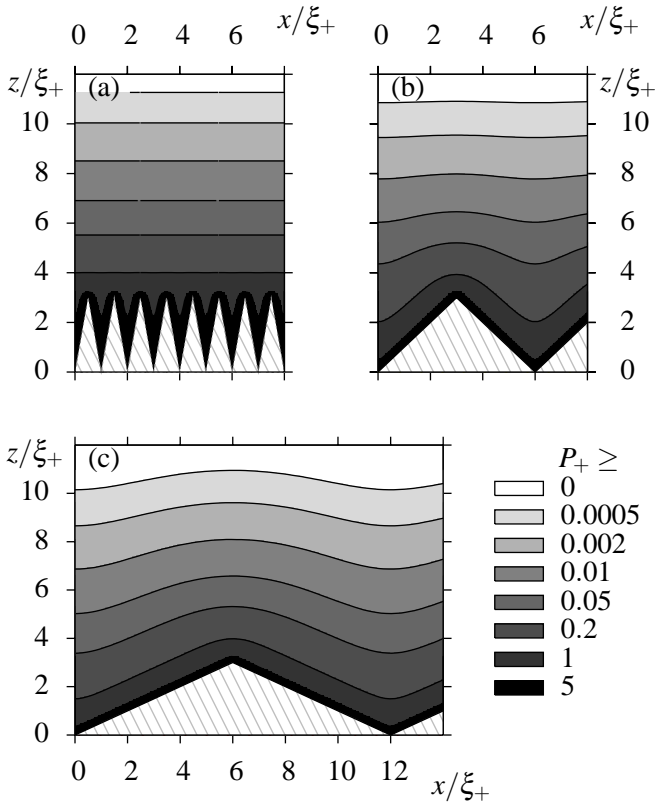}
	\caption{%
		Order parameter scaling function $P_+(x_+,z_+,h_+,\gamma)$ for ${T>T_c}$,
		where $x_+=x/\xi_+$, $z_+=x/\xi_+$, and $h_+=h/\xi_+$,
		in terms of contour plots for geometries with constant scaled corrugation 
		height $h_+=3$ but varying wedge angle $\gamma=$ 
		$19\degree$ in (a), $90\degree$ in (b), and $128\degree$ in (c).
		For small values of $\gamma$ and for increasing normal distances $z_+$
		the profile rapidly adopts an effective
		planar wall behavior corresponding to straight contour lines.
		For large values of $\gamma$ the wall structure reaches deeply
		into the bulk.
		}
	\label{fig:orderparamprofiles}
\end{figure}

In terms of the universal scaling function $P_+$ (see Eq.~\eqref{eq:scaling}) examples of 
order parameter profiles for ${T>T_c}$ are shown in Fig.~\ref{fig:orderparamprofiles}.
(In the case ${T<T_c}$, the scaling function $P_--1$ looks very similar provided $h_-=h_+$
and $\gamma$ are the same.)
%
%
For constant values of $h_\pm$ the influence of the corrugation on the order parameter profile along the
$z$ direction varies strongly with the wedge angle $\gamma$.
For small $\gamma$, upon increasing $z_+$ the profiles rapidly adopt an effective planar wall behavior corresponding
to straight contour lines of the order parameter.
For large wedge angles $\gamma$ the corrugation induces a spatial variation of the
order parameter also far from the wall.
That is, the dependence of the order parameter profile on $\gamma$ is crucial and cannot be
split off in a simple way.
\par
\begin{figure}

	{
	\includegraphics{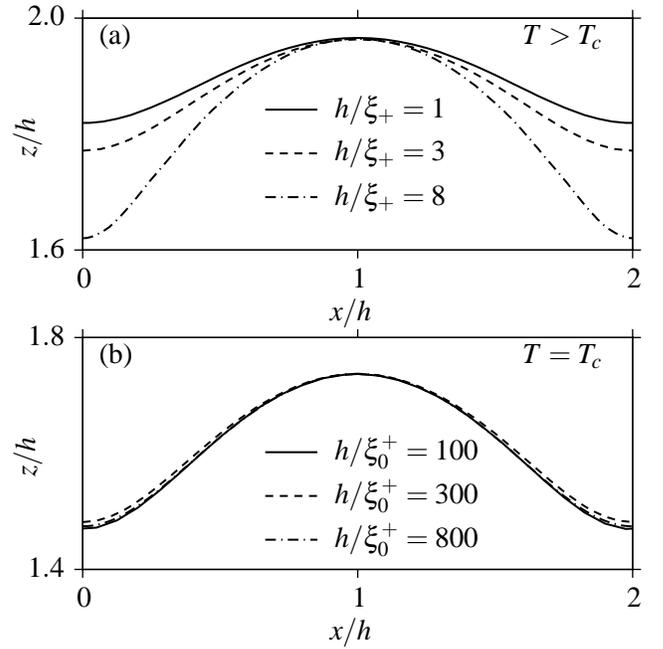}
	}
	\caption{%
		Comparison of the contour lines of the 
		scaling function $P_+$ for ${T>T_c}$ (a) and of the order parameter profile $m$ for ${T=T_c}$ (b),
		respectively,
		for the same wedge angle $\gamma=90\degree$ but for
		various corrugation heights $h$. 
		The values of $P_+$ in (a) and $m$ in (b), respectively, have been
		chosen such that all contour lines coincide at $x/h=1$.
		(The values of $P_+$ in (a) are $1.0$ for $h/\xi_+=1$, 
			$0.093$ for $h/\xi_+=3$, and $0.0005$ for $h/\xi_+=8$;
			the values of $m$ in (b)
			are $1.61\times10^{-2}/\xi_0^+$ for $h/\xi_0^+=100$, $5.3\times10^{-3}/\xi_0^+$ for 
			$h/\xi_0^+=300$, and $2.0\times10^{-3}/\xi_0^+$ for $h/\xi_0^+=800$.)
		For fixed $\gamma$, in (a) the order parameter scaling function for ${T>T_c}$ 
		depends on the values of $z/\xi_+$, $x/\xi_+$, and $h/\xi_+$
		(or equivalently on $z/h$, $x/h$, and $h/\xi_+$),
		in contrast to the profile for ${T=T_c}$ in (b), which
		depends only on the distances $z$ and $x$ scaled by the corrugation height $h$,
		so that for various values of $h$ the contour lines coincide, which
		amounts to a welcome check of the numerical data.
		(The slight deviations of the contour lines 
		in (b) are due to numerical errors in the determination of the contour lines.)
		}
	\label{fig:contourcompare}
\end{figure}
\begin{figure}

	{
	\includegraphics{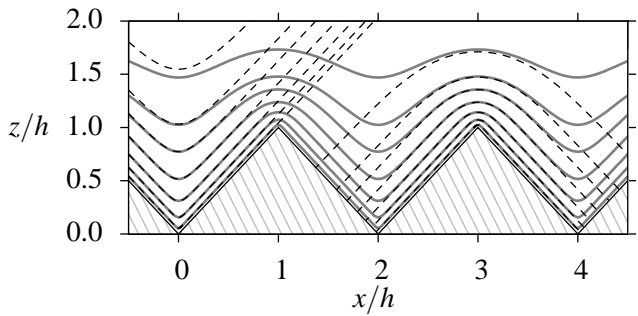}
	}
	\caption{%
		Contour plot of the order parameter profile $m$ at criticality for
		$\gamma=90\degree$.
		The gray solid lines correspond to values of $m\times h=1.6$, $2.4$, $3.2$, $4.8$, $8.0$, $16$, 
		and $48$ (top down).
		For ${T=T_c}$ the value of $m\times h$ depends only on $x/h$, $z/h$, and $\gamma$.
		The corresponding analytic solutions \cite{HANKE99} for single wedges and ridges are 
		shown as dashed lines.
		Close to the edges the order parameter profile for a single wedge or ridge coincides with the
		one close to a periodic array of wedges and ridges.
		}
	\label{fig:contourzero}
\end{figure}

A comparison of contour lines for the order parameter profile at a constant wedge angle $\gamma$ but for varying 
corrugation height is shown in Fig.~\ref{fig:contourcompare} for ${T>T_c}$ and ${T=T_c}$.
For $T= T_c$ the order parameter depends only on the scaled lengths $x/h$ and $z/h$ 
[Fig.~\ref{fig:contourcompare}(b) and Eq.~\eqref{eq:scaling-zero}] in contrast
to the off-critical case, for which the order parameter depends in addition on 
$h_\pm=h/\xi_\pm$ [Fig.~\ref{fig:contourcompare}(a) and Eq.~\eqref{eq:scaling}].
For increasing values of $h_+=h/\xi_+$ the undulation of the contour lines
does not only increase in units of $z/\xi_+$ but also relatively 
in units of $z/h$ as can be inferred from Fig.~\ref{fig:contourcompare}(a).
%
%
In Fig.~\ref{fig:contourzero} a comparison of the order parameter profile for the corrugated
wall geometry with the analytic expression \cite{HANKE99} for the order parameter profile
close to a {single} wedge/ridge is shown.
As expected, close to the edges the analytic solution coincides with the results for the 
geometry of a periodic array of wedges and ridges.
%
%
%
\subsubsection{Proliferation of the surface structure into the bulk \label{sec:undulation}}
\begin{figure}

	\includegraphics{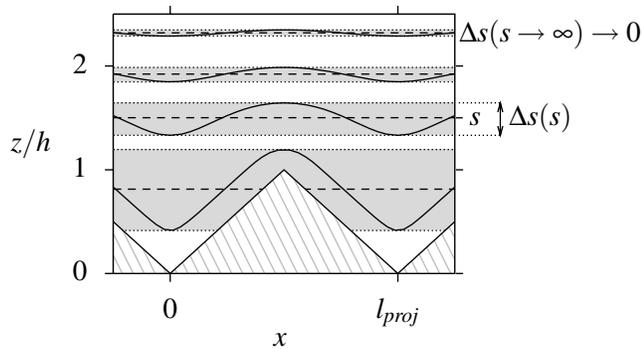}
	\caption{Sketch of the undulation $\Delta s$ of the contour lines 
		of the order parameter profile measured in units of the corrugation height $h$.
		$\Delta s$ depends on the value $s=z/h$ of the mean position of a contour line
		along the $z$ direction.
		The undulation of the contour lines is a measure of how deeply the surface structure reaches
		into the bulk, independent of the absolute value of the order parameter itself.
		For $s\to\infty$ the undulation $\Delta s$ vanishes.
		}
	\label{fig:spatvar}
\end{figure}

%
In order to discuss the propagation of the substrate structure into the bulk in more detail, we
study the behavior of the {contour lines} of the order parameter.
The deviation of a contour line from a straight line gives a measure of how the structure influences
the adsorption behavior at some distance $z$ from the surface, independent of the actual value of the
order parameter at this position.
Using this measure, the cases ${T<T_c}$, ${T=T_c}$, and ${T>T_c}$ can be compared in a straightforward way.
To this end we introduce, in units of $h$, the undulation $\Delta s(s)$ being the width of
a given contour line along the $z$-direction as a function of its corresponding mean position
$s=z/h$ (see Fig.~\ref{fig:spatvar}).
%
%
\par
As a {distant wall} approximation we 
introduce a superposition ansatz for the order parameter,
\begin{equation}
\label{eq:superposition}
	m(x,z,h,\gamma,t)\simeq m^{(0)}(z,t) + m^{(1)}(x,z,h,\gamma,t),
\end{equation}
where $m^{(0)}$ is the effective {planar} wall solution,
and $m^{(1)}$ is the corrugation contribution to the order parameter profile, which is assumed to factorize
into $x$- and $z$-dependent part as
\begin{equation}
\label{eq:product}
	m^{(1)}(x,z,h,\gamma,t)=m^{(x)}(x,h,\gamma,t)\times m^{(z)}(z,h,\gamma,t).
\end{equation}
Inserting the ansatz in Eq.~\eqref{eq:product} into the Euler-Lagrange equation 
[Eq.~\eqref{eq:diffeq}] and using the fact that far from the wall,
where $s\gg 1$ and $\Delta s(s)\ll 1$, the corrugation contribution is much smaller than the planar wall contribution, one finds
for the major corrugation contribution a sinusoidal dependence $m^{(x)}\propto \cos(2\pi x/\lproj)$
along the $x$ direction, and that $m^{(z)}$ is 
decaying \emph{exponentially} along the $z$ direction for all temperatures, including ${T=T_c}$.
This is discussed in detail in Appendix \ref{sec:appendix-1}.
For ${T=T_c}$ this is in contrast to the planar wall contribution, which at $T_c$ decays algebraically.
For $s\gg1$, i.e., within the distant wall approximation, the undulation of the contour lines behaves as
(see Appendix~\ref{sec:appendix-1} or Ref.~~\onlinecite{MT})
\begin{equation}
	\label{eq:delta}
	\Delta s(s)\propto\begin{cases}
			\left(s-\hefftilde\right)^2\exp\{-\sigma_0(\gamma)s\},\;&T=T_c,\\
			\exp\{-\sigma_\pm(h/\xi_\pm,\gamma)s\}\,,&T\lessgtr T_c,
		\end{cases}
\end{equation}
where $\sigma_0$ and $\sigma_\pm$ characterize the exponential decays:
\begin{equation}
	\label{eq:sigma0}
	\sigma_0(\gamma)=\pi\cot(\gamma/2)
\end{equation}
and
\begin{equation}
	\label{eq:sigma}
	\sigma_\pm(h_\pm,\gamma)=h_\pm\left[
		\sqrt{1+\left(\frac{\pi}{h_\pm}\cot(\gamma/2)\right)^2}
		-1\right].
\end{equation}
These expressions are obtained by assuming that the angle $\gamma$ is not too small and,
in the case $t\neq 0$, that $h$ and $\lproj$ are comparable with $\xi_\pm$.
\par
The decay constant $\sigma_0(\gamma)$ for ${t=0}$ corresponds to an exponential decay 
${\propto\exp\{-2\pi z/\lproj\}}$ of the amplitude of the order parameter undulations.
This decay is analogous to substrate potentials near crystal
surfaces with a periodic atomic structure, where 
that part of the potential which is due to the atomic corrugation
decays exponentially, too, although the strength of the substrate potential decays algebraically 
(see, e.g., Refs.~~\onlinecite{STEELE,COLE}).
\par
\begin{figure}

	\includegraphics{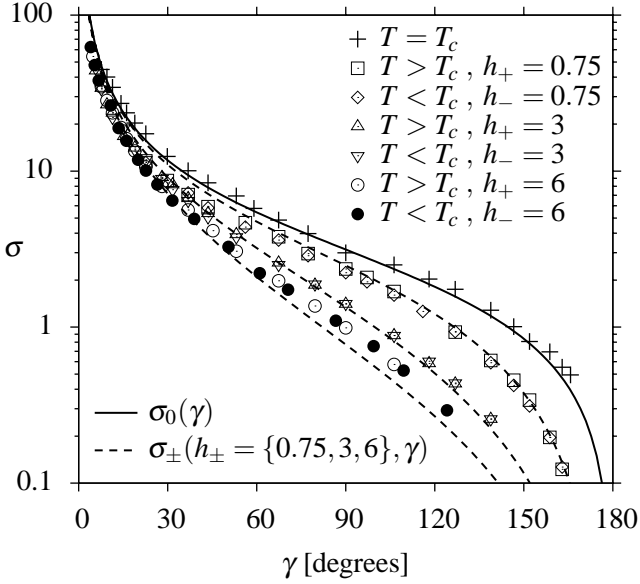}
	\caption{%
		Decay constants $\sigma_0$ (${T=T_c}$) and $\sigma_\pm$ ($T\neq T_c$) corresponding to the 
		exponential decay of the undulation $\Delta s(s)$ (see Fig.~\ref{fig:spatvar}) of the order 
		parameter profiles 
		for critical adsorption on a single patterned substrate (see Fig.~\ref{fig:definitions}).
		The analytic distant wall approximations (lines; Eqs.~\eqref{eq:sigma0} and \eqref{eq:sigma})
		compare well with the full numerical data for the order parameter profiles (symbols).
		There are deviations outside the range of validity 
		of the underlying distant wall approximation.
		Within the numerical accuracy the analytic prediction ${\sigma_+(h_+)}={\sigma_-(h_-)}$
		(Eq.~\eqref{eq:sigma}) is confirmed.
		}
	\label{fig:spatvarcompare}
\end{figure}

In addition, the undulation of the contour lines can be determined from numerically 
calculated order parameter profiles.
The numerical data confirm that the undulations decay exponentially for \emph{all} temperatures.
Moreover, Fig.~\ref{fig:spatvarcompare} shows that the numerically determined decay constants compare well 
with the results obtained within the distant wall approximation (Eqs.~\eqref{eq:sigma0} and \eqref{eq:sigma});
there are significant deviations only outside the range of validity of the approximations
used to derive Eqs.~\eqref{eq:sigma0} and \eqref{eq:sigma}.
%
%
Interestingly, the undulations of the contour lines for ${T=T_c}$ decay more rapidly than for
$T\neq T_c$, in contrast to the order parameter profile itself.
Furthermore, as mentioned above, Eq.~\eqref{eq:sigma} implies that for large values of $h/\xi_\pm$
the effect of structuring the substrate propagates not only in absolute units but also in units of $h$ 
deeper into the bulk than for small values of $h/\xi_\pm$.
\par
For certain substrates it is possible to keep $\lside$, i.e., the actual substrate surface,
constant, but to vary the wedge angle $\gamma$, like an accordion.
Experimentally, such tunable, periodic buckled surfaces can be created using wrinkling polymer films
under stress \cite{Harrison:2004,Genzer:2006}.
For the periodic array of wedges and ridges, the unscaled undulation of the contour 
lines $\Delta z(z)=h\Delta s(z/h)$ is maximal for $\gamma_\Max$ fulfilling
\begin{equation}
	\label{eq:gammamax}
	\left(\frac{\mbox{d}\Delta z(z)}{\mbox{d}\gamma}\right)_{\gamma=\gamma_\Max}=0.
\end{equation}
Within the distant wall approximation given above and for ${T=T_c}$,
upon neglecting all prefactors of the exponential decay of $\Delta s$ Eq.~\eqref{eq:gammamax} 
becomes
\begin{equation}
	\frac{\sin^3(\gamma_\Max/2)}{\cos^2(\gamma_\Max/2)}=\frac{2\pi z}{\lside}.
\end{equation}
This means that the angle $\gamma_\Max$ for maximal undulation depends on ${z/\lside}$, and
for ${z/\lside\to\infty}$ it attains $\pi$,
i.e., if one wants to achieve significant undulations even far away from the substrate its wedges
have to be wide open.
\subsubsection{Distant behavior}
\begin{figure}

	{
	\includegraphics{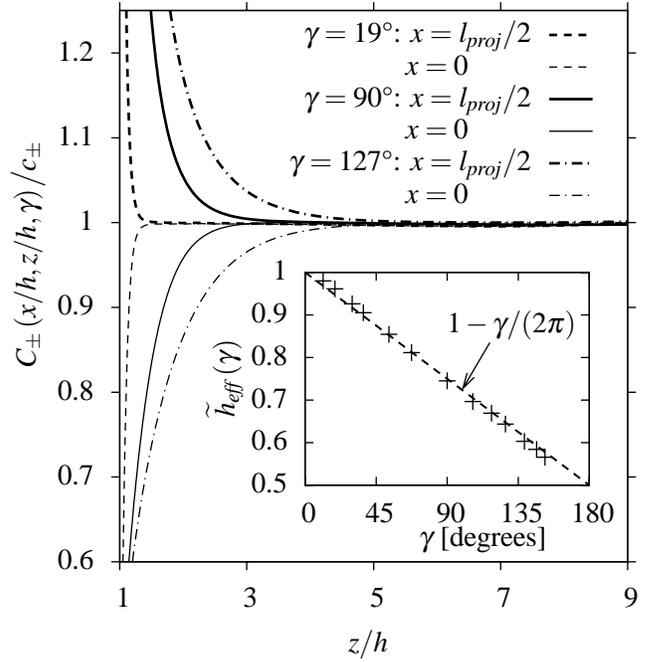}
	}
	\caption{%
		Universal scaling function $C_\pm(x/h,z/h,\gamma)$ for the order parameter at
		${T=T_c}$ [see Eq.~\eqref{eq:definition-function-heff}] in units of the planar wall 
		universal surface amplitude $c_\pm$	[see Eq.~\eqref{eq:definition-heff}].
		The lines represent order parameter profiles corresponding to the lateral position
		of the wedge center ($x=0$) and the edge crest ($x=\lproj/2$)
		for various	values of $\gamma$ [see Fig.~\ref{fig:definitions}].
		Far from the corrugated substrate ($z/h\gg1$) the order parameter resembles the one
		close to a planar wall located the position $z=\heff$.
		The effective substrate height
		$\heff$ is determined such that $C_\pm/c_\pm\to1$ for large $z/h$
		[see Eqs.~\eqref{eq:def-hefftilde}--\eqref{eq:definition-heff}].
		The inset shows the dimensionless effective height $\hefftilde=\heff/h$ (symbols).
		The dashed straight line $1-\gamma/(2\pi)$ approximates the data well.
		}
	\label{fig:heff}
\end{figure}
As shown in Fig.~\ref{fig:heff}, for ${T=T_c}$ the order parameter profile takes the effective 
planar wall form for $z/h\gg1$.
This means that far from the corrugated wall the order parameter resembles the one
for a planar wall located at $z=\heff$.
Based on Eqs.~\eqref{eq:heff-decay} and \eqref{eq:definition-heff} the effective height 
$\hefftilde$ introduced in Eq.~\eqref{eq:def-hefftilde} is determined 
from the order parameter profiles.
From the numerically determined data shown in the inset of Fig.~\ref{fig:heff} we find that
\begin{equation}
	\label{eq:heff-result}
	\hefftilde\simeq 1-\frac{\gamma}{2\pi}.
\end{equation}
For $\gamma\to0$ the effective height is $\heff=h$, which means that the 'closed' wedges form a planar
wall located at $z=h$.
For $\gamma\to\pi$ the effective height attains the value of $\heff=h/2$, and the distant profile 
resembles the one created by a planar wall located at the mean corrugation height.
These limits are consistent with results obtained for the quantum electrodynamic Casimir force 
between corrugated surfaces \cite{EMIG03}.
%
\begin{figure}

	{
	\includegraphics{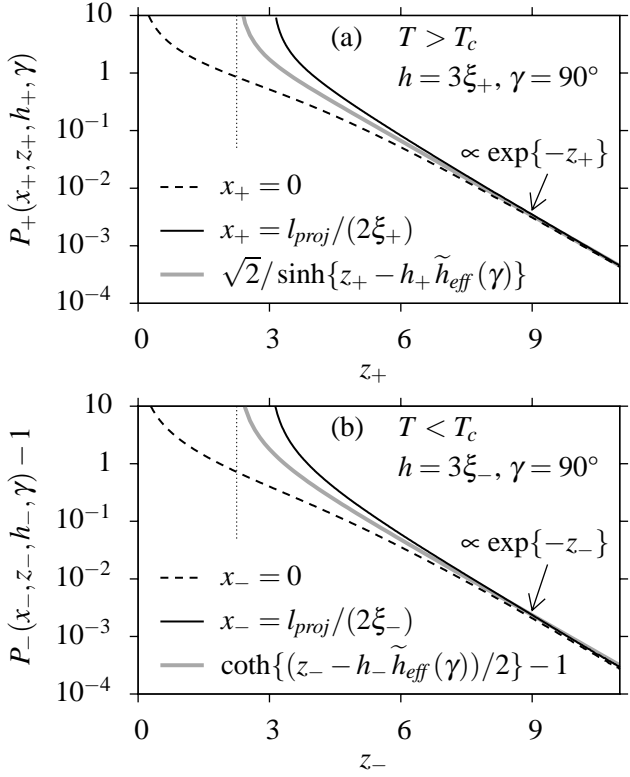}
	}
	\caption{%
		Variation of the order parameter (Eq.~\eqref{eq:scaling}) along the $z$ direction 
		near a periodic array of wedges and ridges (see Fig.~\ref{fig:definitions}) for
		${T>T_c}$ (a) and ${T<T_c}$ (b).
		The corrugation is chosen as $h_\pm=3$ and $\gamma=90\degree$.
		The dashed lines correspond to 
		order parameter profiles at the lateral position $x=0$, i.e., at the center of the wedge,
		the solid lines correspond to the lateral position $x=\lproj/2$, i.e., at the crest of
		the edge (see Fig.~\ref{fig:definitions}).
		For $z/h\gg1$ these profiles join, which means that far from the surface the order 
		parameter exhibits
		an effective planar wall behavior.
		The order parameter decays exponentially towards its bulk value
		$P_+(x_+,z_+\to\infty,h_+,\gamma)=0$, or $P_-(x_-,z_-\to\infty,h_-,\gamma)=1$, respectively.
		The gray lines correspond to the analytic solutions of the
		order parameter close to a planar wall \cite{BINDER83,DIEHL86}
		located at the position $z=\heff=h\,\hefftilde$ (dotted vertical lines) where they diverge.
		The effective substrate height $\heff$ is determined according to Eq.~\eqref{eq:heff-result}.
		It turns out that the value of $\heff$ for a certain wall geometry is the same for
		${T=T_c}$ and $T\neq T_c$.
		}
	\label{fig:cut}
\end{figure}
%
In Fig.~\ref{fig:cut} a cut through the order parameter profile for $T\neq T_c$ is shown.
The order parameter profile exhibits an effective planar
wall behavior with an exponential decay towards its bulk value for $z\gg h$.
It turns out that the effective height for a given wall structure is independent of temperature, i.e.,
the same for ${T=T_c}$ and $T\neq T_c$.
%
%
\subsection{Excess adsorption \label{sec:excess}}
%
Certain experimental approaches such as adsorption measurements provide access only to
integrated quantities rather than spatially resolved local quantities.
For a binary liquid mixture of A and B particles close to its critical 
demixing point at $T_c$ the overall enrichment of, e.g., 
the A particles compared to the B particles 
at a planar wall with perpendicular direction $z$ is given by the excess adsorption per unit wall area
\begin{equation}
	\Gamma^{\infty/2}_\pm(t)=\int_0^\infty dz\,\left\{ \phi^{\infty/2}(z,t)-\phi(z=\infty,t)\right\},
\end{equation}
where the subscripts $\pm$ refer to $T\gtrless T_c$, and
the superscript $\infty/2$ refers to quantities in a semi-infinite system.
The scaling form of the order parameter
does not hold for distances $z$ smaller than or comparable with a typical molecular length $a_0$.
However, for larger distances $z>a_0$ one has \cite{FLOETER95}
\begin{multline}
	\label{eq:flat-excess-adsorption-split}
	\Gamma^{\infty/2}_\pm(t)=%
	\int\limits_0^{a_0}dz\,\left\{ \phi^{\infty/2}(z,t)-\phi(z=\infty,t)\right\}%
	\\
	+a\xi_0^\pm|t|^{\beta-\nu}\int\limits_{a_0/\xi_\pm}^\infty dz_\pm\,\left\{
	P_\pm^{\infty/2}(z_\pm)-P_\pm^{\infty/2}(\infty)\right\}.
\end{multline}
On the molecular scale the order parameter is bounded for $z\to0$, 
whereas the divergence of the universal scaling functions reflects the continuum description.
Thus, the first integral is 
a finite nonuniversal quantity, which is subdominant to the second, diverging integral.
The universal amplitude of the singular part
of the excess adsorption of a planar wall is
\begin{multline}
	\widetilde{\Gamma}_\pm^{\,\infty/2} = 
	\lim_{t\to0}\left[\Gamma_\pm^{\infty/2}(t)\big/\left(\phi(z=\infty,t)\;\xi_\pm\right)\right]
	\\
	 =\int_{0}^\infty dz_\pm\,\left\{ P_\pm^{\,\infty/2}(z_\pm)-P_\pm^{\,\infty/2}(\infty)\right\},
\end{multline}
where the lower limit of the integral has attained zero for $\xi(t\to0)\to\infty$.
\par
For a periodic array of wedges and ridges it is sufficient to consider a single unit cell
as sketched in Fig.~\ref{fig:definitions}.
For one unit cell the excess adsorption $\Gamma_\pm$ per $(d-2)$ dimensional projected area is given by
\begin{equation}
	\label{eq:excess-struc}
	\Gamma_\pm(h,\gamma,t)=a\left(\xi_0^\pm\right)^2|t|^{\beta-2\nu}\,\widetilde{\Gamma}_\pm(h_\pm,\gamma),
\end{equation}
with its scaling function
\begin{multline}
	\label{eq:structure-adsorption}
	\widetilde{\Gamma}_\pm(h_\pm,\gamma)=
	\int_{-\lproj/(2\xi_\pm)}^{\lproj/(2\xi_\pm)}dx_\pm
	\int_{2h_\pm |x|\lproj^{-1}}^{\infty}dz_\pm
	\\
	\left\{ P_\pm(x_\pm,z_\pm,h_\pm,\gamma)-P_\pm^{\infty/2}(z_\pm\to\infty)\right\}
	.
\end{multline}
For a planar substrate, i.e., $\gamma=\pi$, Eq.~\eqref{eq:structure-adsorption} reduces to
\begin{equation}
	\widetilde{\Gamma}_\pm(h_\pm=0,\gamma=\pi)=\frac{\lproj}{\xi_\pm} \widetilde{\Gamma}_\pm^{\,\infty/2}.
\end{equation}
In order to compare the critical adsorption on geometrically structured substrates with 
the critical adsorption on planar substrates, we define a {reduced relative excess adsorption},
\begin{equation}
	\label{eq:upsilon-definition}
	\Upsilon_\pm\left(h_\pm,\gamma,\tfrac{l}{\xi_\pm}\right)
	=\frac{\widetilde{\Gamma}_\pm(h_\pm,\gamma)-\frac{l}{\xi_\pm}\widetilde{\Gamma}_\pm^{\,\infty/2}}
									{\frac{l}{\xi_\pm}\widetilde{\Gamma}_\pm^{\,\infty/2}}.
\end{equation}
$\Upsilon_\pm$ relates the excess adsorption for a unit cell of a corrugated substrate 
to the excess adsorption for a planar substrate of lateral extension $l$ in $x$-direction in units of the latter one.
Obviously, there are two interesting choices for $l$: 
the projected width of a periodic wedge $\lproj$, 
and the actual surface length $\lside$ of a unit wedge.
\par
For these limiting cases one can obtain explicit expressions for $\Upsilon_\pm$.
For $h_\pm=h/\xi_\pm\to\infty$ the corrugation resembles a collection of large tilted planar walls 
of length $\lside$ with a relatively thin layer within which the order parameter
deviates from its bulk value.
For large $\lside$ the edge contributions within a unit cell become unimportant.
Likewise, in the limit $\gamma\to\pi$ the wavelength of the corrugation is very large and
the edge contributions can be neglected, too. 
Thus, in those limits the adsorption in one unit cell of the array of wedges and ridges is the same 
as the adsorption at a planar wall of lateral size $\lside$,
and one has
\begin{equation}
	\label{eq:upsilon-asymp1}
	\Upsilon_\pm\left(h_\pm,\gamma,\tfrac{\lproj}{\xi_\pm}\right)
	\xrightarrow[\mbox{\scriptsize or }\gamma\to\pi]{h_\pm\to\infty}
		\frac{\lside-\lproj}{\lproj}=\frac{1}{\sin\left(\tfrac{\gamma}{2}\right)}-1.
\end{equation}
On the other hand for $h_\pm\to0$ the corrugation becomes vanishingly small on the scale of
the correlation length, and the total adsorption in a unit cell is the same as the one of a 
planar wall of lateral size $\lproj$.
Therefore, one finds
\begin{equation}
	\label{eq:upsilon-asymp2}
	\Upsilon_\pm\left(h_\pm,\gamma,\tfrac{\lside}{\xi_\pm}\right)
	\xrightarrow{h_\pm\to0}
		\frac{\lproj-\lside}{\lside}=\sin\left(\tfrac{\gamma}{2}\right)-1.
\end{equation}
For $\gamma\to0$ and arbitrary values of $h_\pm$ we have not found a simple asymptotic formula.
\par
With Eq.~\eqref{eq:upsilon-definition} we define the ratio $\widetilde{R}$  of the excess adsorption above and below $T_c$ as
\begin{eqnarray}
	\label{eq:R-definition}
	\widetilde{R}(h_+,\gamma)&
	=&
	\frac{\widetilde{\Gamma}_+(h_+,\gamma)}{\widetilde{\Gamma}_-(h_-=h_+,\gamma)}
	\\	&
	=&
	\frac{\xi_-}{\xi_+}\;\frac{\widetilde{\Gamma}_+^{\,\infty/2}}{\widetilde{\Gamma}_-^{\,\infty/2}}
			\;\frac{1+
	\Upsilon_+\left(h_+,\gamma,\tfrac{l}{\xi_+}\right)
			}{1+
	\Upsilon_-\left(h_-=h_+,\gamma,\tfrac{l}{\xi_-}\right)
			},
	\nonumber
\end{eqnarray}
which is independent of the choice of $l$.
$\widetilde{\Gamma}_+^{\,\infty/2}\big/\widetilde{\Gamma}_-^{\,\infty/2}$
is a universal constant for critical adsorption at a planar wall and is experimentally accessible
\cite{FLOETER95}.
\subsubsection{Universal behavior of the excess adsorption within mean field theory}
In the remainder of this subsection we present the results for the excess adsorption obtained from
order parameter profiles calculated within MFT.
Due to the divergence of the order parameter at the surfaces in the limit of strong adsorption, 
which in this scaling limit leads to a divergence of the excess adsorption, 
one has to calculate $\Upsilon_\pm$ carefully.
$\Upsilon_\pm$ is obtained by introducing a cut-off for small distances from the substrate
and subsequent extrapolation for the cut-off sent to zero.
%
\begin{figure}

	\includegraphics{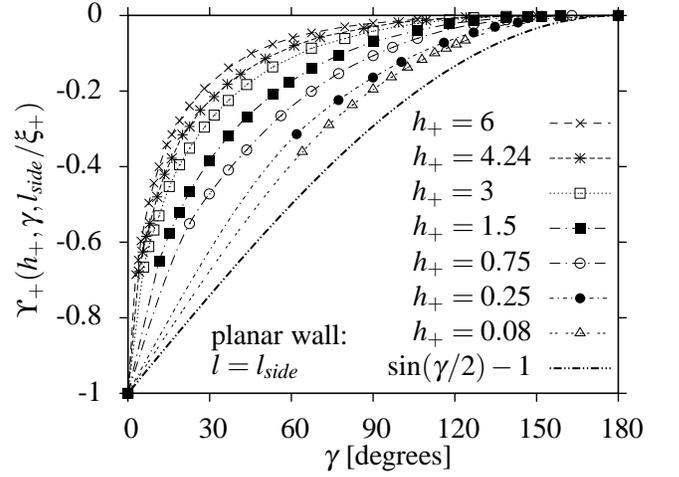}
	\caption{
		Reduced relative excess adsorption $\Upsilon_+$ within MFT 
		at a single geometrically structured substrate [Fig.~\ref{fig:definitions}]
		for ${T>T_c}$ [Eq.~\eqref{eq:upsilon-definition}],
		comparing a structured geometry	with a planar wall of lateral size $\lside$.
		Note that $\lside=2h/\cos(\gamma/2)$.
		The lines connecting data points are smooth fits, while the thick
		dashed-double-dotted line shows the asymptotic behavior for $h_+\to0$ [Eq. \eqref{eq:upsilon-asymp2}].
		The negative range of values of $\Upsilon_+$ means that the amount of adsorbed matter on a 
		geometrically structured substrate is less than the one on a planar wall with the 
		same actual substrate surface size.
		All curves meet at the point $\{\gamma=0,\Upsilon_+(h_+,\gamma,\lside/\xi_+)=-1\}$, because
		for 'closed' wedges the excess adsorption in a unit cell of the periodic array of wedges and
		ridges vanishes compared with the reference value of the excess adsorption 
		on a planar substrate with the same actual surface size.
		}
	\label{fig:side}
\end{figure}
\begin{figure}

	\includegraphics{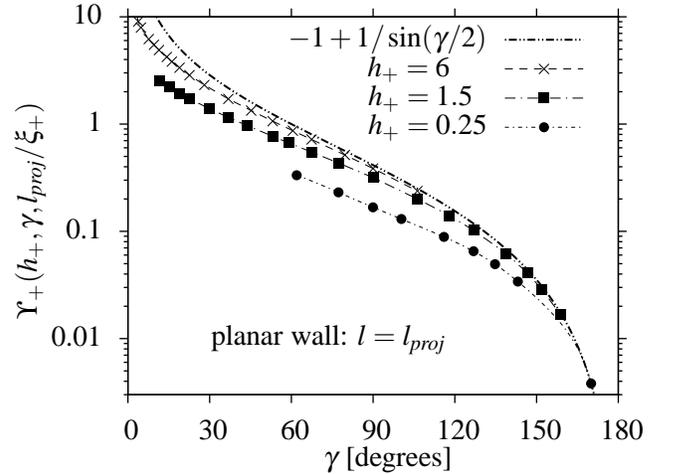}
	\caption{
		Reduced relative excess adsorption $\Upsilon_+$ within MFT at a corrugated substrate 
		[Fig.~\ref{fig:definitions}]
		for ${T>T_c}$ [Eq.~\eqref{eq:upsilon-definition}]
		in comparison with the one on a flat surface of lateral size $\lproj$.
		Note that $\lproj=2h\;\tan(\gamma/2)$.
		The lines connecting data points are smoothly fitted, whereas the thick
		dashed-double-dotted line shows the asymptotic behavior for $\gamma\to\pi$ or $h_+\to\infty$ 
		[Eq. \eqref{eq:upsilon-asymp1}].
		The excess adsorption on a geometrically structured substrate is larger than the one
		on a planar substrate of the same projected area.
		For $\gamma\to0$ the ratio of these two quantities even diverges.
		This is in contrast to the case $l=\lside$ in Fig.~\ref{fig:side}.
	}
	\label{fig:proj}
\end{figure}
%
\par
Figure~\ref{fig:side} displays the reduced relative excess adsorption for $t>0$ in comparison
to a planar wall of lateral size $l=\lside$; in Fig.~\ref{fig:proj} the comparison with a planar 
wall of lateral size $l=\lproj$ is shown.
The general asymptotic behaviors derived above (Eqs.~\eqref{eq:upsilon-asymp1} and \eqref{eq:upsilon-asymp2})
are attained in all cases.
Concerning the interesting behavior for small wedge angle $\gamma$, 
the total excess adsorption is less than the one for planar walls of lateral size $\lside$.
This means that for all corrugation parameters $h_\pm$ and $\gamma$
the loss in adsorption near an edge exceeds the gain in adsorption inside a wedge \cite{PALAGYI04};
this effect becomes more pronounced upon approaching $T_c$.
However, for $\gamma\to0$ the amount of adsorbed matter is larger than at a wall of lateral size $\lproj$.
Indeed, $\Upsilon_\pm(h/\xi_\pm,\gamma,\lproj/\xi_\pm)$ even seems to diverge for $\gamma\to0$, i.e., 
the adsorption on a geometrically structured substrate with a short corrugation wavelength is 
much larger than the one for a planar substrate with the same projected area.
This effect becomes more pronounced away from $T_c$.
Thus, if one aims at enhancing the effect of critical adsorption, one should structure the
adsorbing substrate.
At a first glance this appears to contradict the conclusion above that the order parameter 
profile for $z>h$ and for $\gamma\to0$ closely resembles that of a planar wall 
[Fig.~\ref{fig:orderparamprofiles}(a)].
However, for $T\neq T_c$ and within MFT the major contribution to the adsorption 
stems from the vicinity of the substrate.
This part diverges due to the increase $\propto R^{-1}$ of the order parameter profile
near the surface, where $R$ is the closest distance to the surface.
That is, within MFT the major contribution to the adsorption is governed by the wedge walls.
Since the wedge walls are accessible for the fluid even for $\gamma\to0$ and because the size of the wedge
walls is larger than $\lproj$, the relative excess adsorption diverges for $\gamma\to0$ 
(which implies $\lproj\to0$).
However, 
in order that the actual physical quantities are described properly by the universal scaling functions,
$\lproj$ must be large compared with molecular length scales, which imposes a limit on
the smallest wedge angle below which the universal features do no longer capture the structure
near the wedge center.
\par
Beyond MFT, the behavior of $\Upsilon_\pm(h_\pm,\gamma,\lproj/\xi_\pm)$ is expected to be 
qualitatively different.
In this case the critical exponent ratio 
for the order parameter decay at $T_c$ and at a planar substrate (Eq.~\eqref{eq:scaling-zero})
is $\beta/\nu\simeq0.5$
(instead of $\beta/\nu=1$ within MFT), and the divergence of
the excess adsorption stems from the slow decay \emph{far} from the surface and occurs only for 
$T\to T_c$.
Beyond MFT, there is no diverging contribution from the vicinity of the substrate.
Since for small $\gamma$ the order parameter profile outside the wedges of the periodic array
of wedges and ridges resembles the one of a planar wall (see Fig.~\ref{fig:orderparamprofiles}(a)),
one expects the excess adsorption in a unit cell of the periodic array of wedges and ridges to
approach for $\gamma\to0$ the value of the excess adsorption at a planar wall of lateral size
$\lproj$.
That is, $\Upsilon_\pm(h_\pm,\gamma\to0,\lproj/\xi_\pm)\to0$ beyond MFT (in contrast to the divergence
within MFT [Fig.~\ref{fig:proj}]).
However, for intermediate wedge angle ranges $0<\gamma<\pi$ the excess adsorption at a corrugated substrate
is larger than at a flat one, i.e., generally $\Upsilon_\pm(h_\pm,\gamma,\lproj/\xi_\pm)>0$.
On the other hand the behavior of $\Upsilon_\pm(h_\pm,\gamma,\lside/\xi_\pm)$, which measures the
difference of the excess adsorption at a corrugated substrate and a planar wall of lateral extension
$\lside$, should be qualitatively similar to its behavior within MFT [Fig.~\ref{fig:side}] because
also within MFT a major contribution to this difference is due to behavior distant from the walls.
Beyond MFT, due to the slow decay of the order parameter far from the wall for $T\to T_c$, 
the property of the excess adsorption at a corrugated wall to be smaller than the corresponding 
one at a planar wall of lateral extension $\lside$ is even more pronounced than within MFT.
That is, also beyond MFT $\Upsilon_\pm(h_\pm,\gamma,\lside/\xi_\pm)<0$, and 
$\Upsilon_\pm(h_\pm,\gamma\to0,\lside/\xi_\pm)\to-1$.
\par
\begin{figure}

	\includegraphics{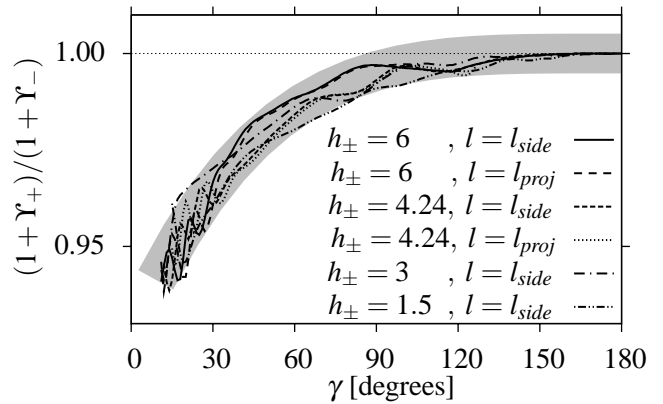}
	\caption{Universal ratio $\widetilde{R}$ of the excess adsorptions 
			at a single topologically structured substrate [see Fig.~\ref{fig:definitions}]
			above and below $T_c$ [Eq.~\eqref{eq:R-definition}],
			divided by the corresponding planar wall ratio and $\xi_-/\xi_+$.
			The scatter of the data is due to 
			the difficulty to interpolate between the data points.
			Besides these uncertainties the curves 
			for different $h_+=h_-$ and $l=\lside,\lproj$ do not differ much.
			Approximately, for all parameters the data vary as a function of $\gamma$ as
			$\simeq 1-0.07[(\gamma-\pi)/\pi]^4$ (shaded region).
			The ratio of the excess adsorptions above and below $T_c$ is smaller at rough substrates than
			on planar ones,	especially for small values of $\gamma$.
			}
	\label{fig:ratio}
\end{figure}

For $t<0$ the reduced relative excess adsorption looks similar to the case $t>0$.
However, the 
curves for $h_-=h_+$ do {not} fall on top of each other.
One can determine the ratio $\widetilde{R}$ of the excess adsorption above and below $T_c$ as introduced in
Eq.~\eqref{eq:R-definition} by studying these differences.
The dependence of $\widetilde{R}$ on various parameters is shown in Fig.~\ref{fig:ratio}.
%
%
According to its definition in Eq.~\eqref{eq:R-definition}, $\widetilde{R}$ is 
independent of the choice for $l$.
In addition, it seems to be independent of the choice for $h_+$, too.
$\widetilde{R}$ attains $\frac{\xi_-}{\xi_+}\;\frac{\widetilde{\Gamma}_+^{\infty/2}}{\widetilde{\Gamma}_-^{\infty/2}}$ for $\gamma\to\pi$, but decreases for $\gamma\to0$.
Thus, the ratio of the excess adsorption on corrugated substrates above and below $T_c$ 
is different from the corresponding planar wall ratio.
%
%
\section{Critical Casimir forces between geometrically structured substrates}
\label{sec:casimir}
In this section we discuss critical Casimir forces mediated by
critical fluids confined between {two} opposing geometrically structured walls.
Two types of systems are considered: 
First, we study the {normal} critical Casimir force between
a geometrically structured wall and a planar wall. Second, we discuss the {lateral} and the normal
critical Casimir force between two identically structured walls.
Throughout this section 
the chemical boundary conditions are the same for the two walls, i.e., the signs of the surface
fields at the two substrates are equal.
As before, the geometric structure of the substrates is taken to be
a periodic array of wedges and ridges along the same lateral direction $x$.
The corrugations are on top of walls separated by a distance $L$ in $z$-direction
(see Figs.~\ref{fig:defstrucflat} and, c.f., \ref{fig:defstrucstruc}).
As before, the projected $(d-1)$ dimensional area $A$ of the two substrates within
the $\{x,\vec{y}\}$ plane is macroscopically large.
If both substrates are structured, the wedges and ridges of the two substrates can be shifted by $D$ 
relative to each other.
Thus, both cases, a structured substrate opposing a planar one, or two identically structured substrates 
opposing each other, are described by the variables $L$, $h$, and $\gamma$ (and $D$ for the latter case),
in addition to the reduced temperature $t$ as a thermodynamic variable.
%
\subsection{Stress tensor in periodically structured confinements \label{sec:stress}}
\label{sec:forces}
The free energy of such a confined fluid system decomposes into four distinct contributions 
\cite{Schoen:2007}:
\begin{equation}
	\label{eq:finitefree}
	F= V F_b + S F_s + E F_e + \delta F,
\end{equation}
where $V$ is the volume accessible to the fluid,
$S$ denotes the actual substrate surface area, $E$ is the sum of the edge lengths,
and $F_{b,s,e}$ are the bulk, surface, and edge free energy densities.
$\delta F$ corresponds to that part of the free energy stemming from
the finite size effect and the effect of corrugation.
In the following we focus on the singular part of $F$ near the \emph{bulk} critical point
(without introducing a separate notation for this part of $F$) \cite{DIEHL86,CARDY83}.
\par
The {normal} critical Casimir force acting on the confining walls is defined as
\begin{equation}
	\label{eq:fnormaldef}
	f_\perp
	=
	-\frac{1}{k_BTA}\frac{\partial}{\partial L}
	\delta F,
\end{equation}
which does not include the bulk contribution to the force $-\frac{F_b}{k_BTA}\;\frac{\partial V}{\partial L}$,
which is constant with respect to $L$,
and the {lateral} critical Casimir force is
\begin{equation}
	\label{eq:flateraldef}
	f_\parallel
	=
	-\frac{1}{k_BTA}\frac{\partial}{\partial D}
	\delta F.
\end{equation}
\par
By using the stress tensor, the forces can be calculated directly from the
order parameter profiles.
This has the advantage that one does not face the numerical difficulties of calculating differences of free
energies which diverge due to the divergence of the order parameter profiles in the scaling limit
near the surface.
For the fixed point Hamiltonian given by Eqs.~\eqref{eq:hamiltonian} and \eqref{eq:bulkhamiltonian}
the stress tensor components are \cite{COLLINS76,BROWN80,EISENRIEGLER94,KRECH97}
\begin{equation}
	\label{eq:stresstensor}
	T_{kl}(\vec{r})=\frac{\partial \phi}{\partial r_k}\frac{\partial \phi}{\partial r_l}
					-\delta_{k,l}\left[\frac{1}{2}\left(\nabla\phi\right)^2+\frac{\tau}{2}\phi^2
					+\frac{u}{4!}\phi^4\right]
					-I_{kl}(\vec{r}),
\end{equation}
where $k,l=x,z,\ldots$ indicate the components of the position vector 
$\vec{r}=\{x,z,\ldots\}$, and
$I_{kl}$ is the improvement term, which reads
\begin{equation}
	\label{eq:improvementterm}
	I_{kl}(\vec{r})=\frac{d-2}{4(d-1)}\left[\frac{\partial^2\phi^2}{\partial r_k \partial r_l}
					-\delta_{k,l}\nabla^2\phi^2\right].
\end{equation}
For the systems under consideration, $T_{zz}$ and  $T_{xz}$ correspond to the normal
and lateral critical Casimir forces, respectively, as introduced above if bulk
contributions are subtracted (within MFT these occur only for ${T<T_c}$).
\par
In the following we only  consider confinements with {overall} periodicity.
This is the case for the corrugation model used here with an overall periodicity wavelength $\lproj$
if the ratio of the two single corrugation wavelengths is a rational number or if one of the 
two substrates is flat. 
Then the force in $k$-direction per unit area is
\begin{equation}
	\label{eq:stresstensor-2}
	f_k=\frac{1}{\lproj}\int_0^{\lproj}\,dx\,T_{kz},
\end{equation}
where $T_{kz}$ is evaluated at a fixed position $z=z_0$ in between the two substrates.
Due to the periodicity of the system, the contribution to the force stemming from
the improvement term [Eq.~\eqref{eq:improvementterm}] vanishes \cite{MT}:
\begin{equation}
	\frac{1}{\lproj}\int_0^{\lproj}\,dx\,I_{kz}=0.
\end{equation}
%
\subsection{Normal critical Casimir force between a geometrically structured and a flat substrate \label{sec:normal}}
%
\begin{figure}

	\includegraphics{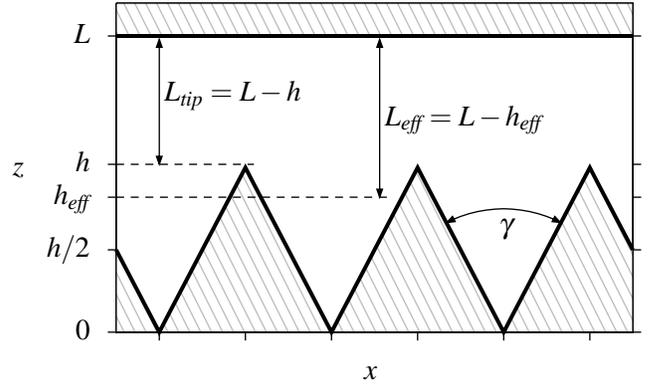}
	\caption{Confinement composed of a \emph{s}tructured wall and a \emph{f}lat wall 
	(denoted as $\Sf$ geometry).
		The wedge vertices of the geometrically structured substrate are separated
		by the distance $L$ from the planar wall.
	}
	\label{fig:defstrucflat}
\end{figure}
%
As a first step for studying the effect of corrugation on critical Casimir forces,
this subsection deals with a \emph{s}tructured wall opposing a \emph{f}lat wall as shown in 
Fig.~\ref{fig:defstrucflat}, denoted as $\Sf$ geometry.
Within the $\Sf$  geometry there is no lateral force.
The normal critical Casimir force takes the following scaling form:
\begin{align}
	\label{eq:crit-force-sf-1}
	f_{\perp}^\Sf(L,h,\gamma,t) &=\; 
				(d-1) \; L^{-d}\;
					\widetilde{f}_{\perp,\pm}^{\,\,\Sf}(y_\pm,h_\pm,\gamma)
					\\
				&\xrightarrow{T\to T_c}\; 
					(d-1) \; L^{-d}\;\Delta_{+,+}\,
					\widetilde{\Delta}_{\perp}^\Sf\left(\tfrac{L}{h},\gamma\right),\nonumber
\end{align}
where $y_\pm=t (L/\xi_0^\pm)^{1/\nu}$, and
$h_\pm=h/\xi_\pm$;
$\widetilde{f}_{\perp,\pm}^{\,\,\Sf}$ is the universal scaling function
of the normal force, 
$\widetilde{\Delta}_{\perp}^\Sf$ is the universal generalized Casimir
amplitude,
and $\Delta_{+,+}$ is the universal Casimir amplitude for two planar plates with parallel surface fields
\cite{KRECH94,Vasilyev:2007}.
For MFT $d=4$, and $y_+=\tau L^2$ for ${T>T_c}$, and $y_-=2\,\tau L^2$ for ${T<T_c}$; the Casimir amplitude
is $\Delta_{+,+}\simeq-15.7561\,(3!/u)$ \cite{KRECH97}.
For ${T=T_c}$ the dimensionless quantity $f_{\perp}^\Sf\times h^d$ depends on $L/h$ and $\gamma$, only 
[Eq.~\eqref{eq:crit-force-sf-1}].
\subsubsection{Distant behavior}
The large distance behavior of the normal force is governed by the {effective} planar wall behavior of the 
corrugated substrate. 
That is, the scaling functions of the normal critical Casimir force resemble the one for 
two planar walls at a distance $\Leff=L-\heff$ (see Eq.~\eqref{eq:crit-force-sf-1}):
\begin{equation}
	\label{eq:force-sf-eff}
	\widetilde{f}_{\perp,\pm}^{\,\,\Sf}(y_\pm,h_\pm,\gamma)
	\xrightarrow{L\gg h,\lproj}
	\left(\frac{L}{\Leff}\right)^d\widetilde{f}_{\perp,\pm}^{+,+}
						\left(y^\pm_{\textit{eff}}\right)
\end{equation}
and
\begin{equation}
	\label{eq:crit-force-sf-eff}
	\widetilde{\Delta}^\Sf_{\perp}\left(\frac{L}{h},\gamma\right)
	\xrightarrow{L\gg h,\lproj}
	\left(1-\left[\frac{L}{h}\right]^{-1}\;\hefftilde\right)^{-d},
\end{equation}
where $y^\pm_{\textit{eff}}=t (\Leff/\xi_0^\pm)^{1/\nu}$ 
and $\widetilde{f}_{\perp,\pm}^{+,+}(y^\pm_{\textit{eff}})$ is the 
scaling function 
for the critical Casimir force between two {planar} walls at distance $\Leff$ 
\cite{KRECH91,KRECH92a,KRECH92b,KRECH94,KRECH97,Vasilyev:2007}.
(In principle $\heff=h\;\hefftilde$ for the confined system 
could differ from the one introduced in Eq.~\eqref{eq:def-hefftilde} for the corresponding 
semi-infinite system; however, it turns out that they are the same.)
%
\begin{figure}

	\includegraphics{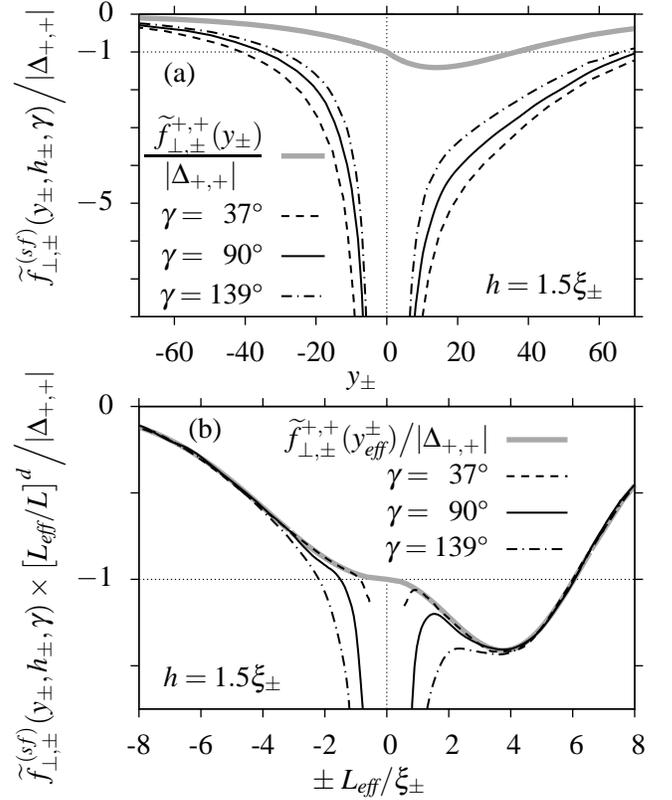}
	\caption{%
		(a) Universal scaling function of the normal critical Casimir force	(in units of the absolute 
		value of the universal critical Casimir amplitude $\Delta_{+,+}$ for planar walls)
		for the $\Sf$ geometry [Fig.~\ref{fig:defstrucflat}] within MFT, for constant
		$h_\pm=1.5$ and for three wedge angles $\gamma$, as a function of $y_\pm=t(L/\xi_0^\pm)^{1/\nu}$.
		The gray line corresponds to the MFT result for the normal critical Casimir force 
		between two planar walls at distance $L$ \cite{KRECH97}.
		The absolute value of the force between a geometrically structured substrate and 
		a planar wall at distance $L$ [Fig.~\ref{fig:defstrucflat}] is stronger compared
		to the one between two planar walls at distance $L$,
		because the corrugation on top of the lower substrate in Fig.~\ref{fig:defstrucflat}
		effectively decreases the distance between the two opposing substrates.
		The geometric constraint $L>h$ [Fig.~\ref{fig:defstrucflat}] implies $|y_\pm|>h_\pm^{1/\nu}$
		and thus leads, for $h_\pm$ fixed, to a divergence of the scaling function 
		of the normal critical Casimir force for
		$|y_\pm|\searrow|y_{h\pm}|\equiv|t|(h/\xi_0^\pm)^{1/\nu}$, 
		where the planar wall approaches the tips of the corrugation; 
		for fixed $h_\pm$ smaller absolute values of $y_\pm$
		cannot be reached for a given $\Sf$ geometry.
		From Eq.~\eqref{eq:derjaguin-2b} one infers that, for fixed $h_\pm\neq0$,
		$\widetilde{f}^{\,\,\Sf}_{\perp,\pm}(y_\pm\to y_{h\pm},h_\pm,\gamma)\propto
		\left||y_\pm|-|y_{h\pm}|\right|^{-\nu(d-1)}$.
		On the other hand, for $t\to0$ both $h_\pm$ and $y_\pm$ vanish so that in this limit the scaling
		variable $y_\pm$ can reach zero, leading to the finite result for $\widetilde{f}_{\perp,\pm}^{\,\,\Sf}$
		given by the second line of Eq.~\eqref{eq:crit-force-sf-1}.
		(b) Rescaled (i.e., multiplied by $(\Leff/L)^d$; see Eqs.~\eqref{eq:force-sf-eff} and 
		\eqref{eq:crit-force-sf-eff}) universal scaling function of the normal critical Casimir force 
		within MFT ($d=4$) for the same confinements as in (a) as a function of $\pm \Leff/\xi_\pm$, 
		where $\pm$ indicates $T\gtrless T_c$.
		Note that the curve for $\gamma=37\degree$ diverges to $-\infty$, too; however,
		due to numerical limitations, here only the indicated limited range is covered.
		The gray line corresponds to the normal critical Casimir force between two planar walls 
		at distance $\Leff$ (see Fig.~\ref{fig:defstrucflat} and the main text); 
		note that within
		MFT $y_{\textit{eff}}^+=\tau \Leff^2$, and $y_{\textit{eff}}^-=2\tau \Leff^2$.
		For $L/h\gg1$ the normal critical Casimir force between a corrugated and a flat substrate
		reaches this effective planar wall limit [Eqs.~\eqref{eq:force-sf-eff} and 
		\eqref{eq:crit-force-sf-eff}] with $\Leff=L-\heff$ and $\heff=h\,\hefftilde$
		given by the semi-infinite expression for $\hefftilde$
		[inset of Fig.~\ref{fig:heff}].
		}
	\label{fig:casimir-normal-distant}
\end{figure}
%
%
\par
In order to provide quantitative expressions for the universal scaling functions we calculate the
stress tensor components within MFT
(see Subsec.~\ref{sec:opm}) corresponding to the zeroth order of the $\varepsilon$-expansion.
These numerical results for the scaling functions of the normal force are shown in 
Fig.~\ref{fig:casimir-normal-distant}, where
the effective height is the same as the one for a single corrugated substrate [Eq.~\eqref{eq:heff-result}].
The data agree very well with the effective planar wall limit for large values of $L/\xi_\pm$.
We find that for constant $h_\pm$ the effective planar wall limit of the normal force
is reached already for smaller values of $L/\xi_\pm$ if $\gamma$ is smaller 
(see Fig.~\ref{fig:casimir-normal-distant}(b)).
On the other hand, we find that for constant $\gamma$ the effective planar wall limit is reached 
faster as a function of $L/\xi_\pm$ if
the value of $h_\pm$ is smaller.
%
\begin{figure}

	\includegraphics{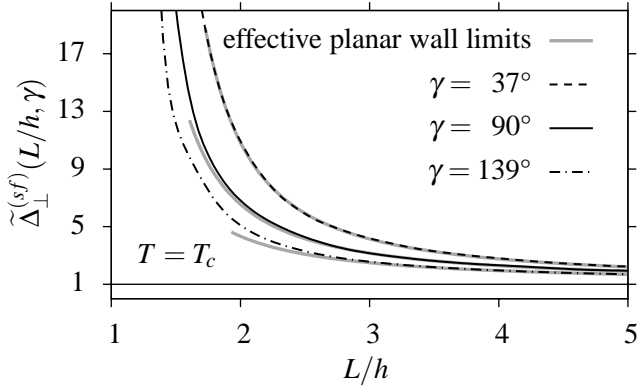}
	\caption{%
		Universal generalized Casimir amplitude $\widetilde{\Delta}_\perp^\Sf(L/h,\gamma)$
		[Eq.~\eqref{eq:crit-force-sf-1}]
		of the normal critical Casimir force at criticality
		for the $\Sf$ geometry
		(see Fig.~\ref{fig:defstrucflat})
		as a function of $L/h$. The results correspond to MFT ($d=4$).
		In this representation the critical Casimir force between two planar walls 
		at distance $L$ corresponds to unity (horizontal line).
		For $L/h\gg1$ the normal critical Casimir force at ${T=T_c}$ approaches
		the corresponding effective planar wall limit [Eq.~\eqref{eq:crit-force-sf-eff}]
		with $\Leff=L-h\,\hefftilde$ 
		where $\hefftilde$ is given by the semi-infinite expression
		[inset in Fig.~\ref{fig:heff}].
		The effective planar wall limits would diverge at $L/h=\hefftilde\le1$
		as $(L/h-\hefftilde)^{-d}$.
		However, this approximation becomes unreliable for small distances.
		The actual scaling function $\widetilde{\Delta}_\perp^\Sf$ diverges at $L/h=1$ as
		$(L/h-1)^{-(d-1)}$ (see Eq.~\eqref{eq:derjaguin-2}).
		For wide wedge angles $\gamma$ the deviations from the effective planar wall limit are most
		pronounced.
		}
	\label{fig:casimir-zero}
\end{figure}
%
Quantitative results for the generalized Casimir amplitude $\widetilde{\Delta}_\perp^\Sf$
[Eq.~\eqref{eq:crit-force-sf-1}] are presented in Fig.~\ref{fig:casimir-zero}.
The numerical data agree with the effective height limit 
(Eq.~\eqref{eq:crit-force-sf-eff}  together with the semi-infinite result for 
$\hefftilde$ in Eq.~\eqref{eq:heff-result})
for large values of $L/h$.
The expression for the effective height [Eq.~\eqref{eq:heff-result}] is consistent with
results for the quantum electrodynamic Casimir force between a corrugated and a planar wall \cite{EMIG03}.
\subsubsection{Nearby behavior}
%
As indicated in Figs.~\ref{fig:casimir-normal-distant} and \ref{fig:casimir-zero} there are 
pronounced deviations
from the effective planar wall limit if the wall distance is small, i.e., for $\Ltip=L-h\to0$.
If the characteristic sizes of the corrugation are much larger than the distance $\Ltip$, a Derjaguin-like approximation may be appropriate 
\cite{DERJAGUIN34}.
Within this Derjaguin-like approximation we replace the corrugation of wedges and ridges by a set of 
staircases with infinitely small horizontal terraces and vertical steps
and sum the single contributions of the terraces in order to obtain the total normal force:
\begin{equation}
	\label{eq:derjaguin-1}
	f_\perp^{(Der)}=
		\frac{2}{\lproj}
		\int\limits_0^{\lproj/2}dx\,
			\frac{d-1}{\left(L(x)\right)^d}\,
			\widetilde{f}_{\perp,\pm}^{+,+}\left((L(x)/\xi_\pm)^{1/\nu}\right)
			,
\end{equation}
where $L(x)=\Ltip+h(1-2|x|/\lproj)$ is the local wall distance and $\widetilde{f}_{\perp,\pm}^{+,+}(y_\pm(x))$
with $y_\pm(x)=(L(x)/\xi_\pm)^{1/\nu}$ is the normal critical Casimir force scaling function between two
planar walls at distance $L(x)$.
The summation of single contributions is clearly a strong approximation to the actual non-local character of
critical Casimir forces and it is reasonable only in the nearby range where local force contributions
govern the behavior.
Performing an integration by parts in Eq.~\eqref{eq:derjaguin-1} one finds that
\begin{multline}
	f_\perp^{(Der)}=
		\frac{1}{h}\left\{
			\frac{1}{\Ltip^{d-1}}
			\widetilde{f}_{\perp,\pm}^{+,+}\left((\Ltip/\xi_\pm)^{1/\nu}\right)\right.
			\\
			\left.-
			\frac{1}{\left(\Ltip+h\right)^{d-1}}
			\widetilde{f}_{\perp,\pm}^{+,+}\left((\Ltip/\xi_\pm+h_\pm)^{1/\nu}\right)
		\right\}
		\\
		+
		\frac{1}{h}\int\limits^{\Ltip+h}_{\Ltip}dL\,
			\frac{(L/\xi_\pm)^{(1-\nu)/\nu}}{\nu\,L^{d-1}}\,
			\widetilde{f'}_{\perp,\pm}^{+,+}\left((L/\xi_\pm)^{1/\nu}\right)
			,
\end{multline}
where $\widetilde{f'}_{\perp,\pm}^{+,+}(y_\pm)=\frac{d}{dy_\pm}\widetilde{f}_{\perp,\pm}^{+,+}(y_\pm)$.
Within MFT the absolute values of $\widetilde{f}_{\perp,\pm}^{+,+}(y_\pm)$ and its derivatives with respect 
to $y_\pm$ do not diverge and are generally decreasing with increasing $|y_\pm|$, or at least they are not 
much larger than their corresponding values for smaller absolute values of $y_\pm$ \cite{KRECH97}.
Thus, in the limit $\Ltip/h\ll1$ one finds for the \emph{leading} behavior of the critical Casimir force 
\begin{equation} 
	\label{eq:derjaguin-2b}
	f_\perp^{(Der)}\simeq
	\frac{1}{h}\Ltip^{-(d-1)}\,
	\widetilde{f}_{\perp,\pm}^{+,+}\left((\Ltip/\xi_\pm)^{1/\nu}\right).
\end{equation}
That is, at short distances the normal critical Casimir force obeys an algebraic behavior 
$\propto \Ltip^{-(d-1)}$ with a power law distinct from that for the force between two planar walls.
For $\Ltip/\xi_\pm\to0$ (or $T\to T_c$, respectively) one finds 
\begin{equation}
	\label{eq:derjaguin-2}
	f_\perp^{(Der)}\simeq\frac{\Delta_{+,+}}{h}\Ltip^{-(d-1)}.
\end{equation}
Beyond MFT, the general trends of the scaling function for the critical Casimir force between planar walls
are similar to those within MFT \cite{Hucht:2007,Vasilyev:2007}.
Therefore, we expect Eq.~\eqref{eq:derjaguin-2b} to be a reliable approximation in the limit $\Ltip/h\to0$ 
also beyond MFT.
\par
\begin{figure}

	\includegraphics{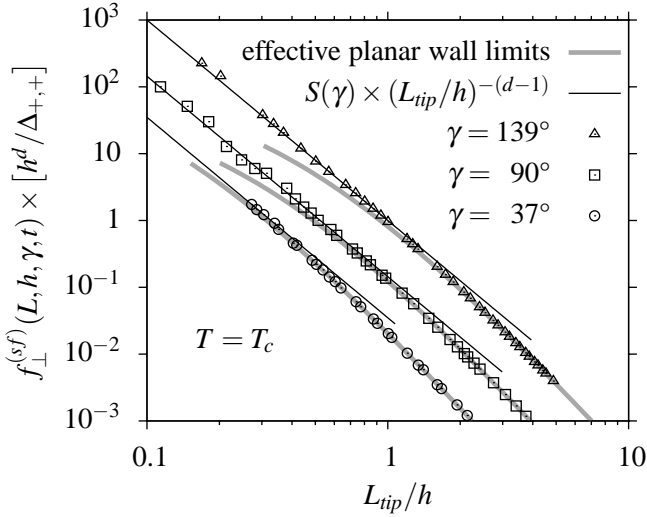}
	\caption{%
		Normal critical Casimir force $f_\perp^\Sf(L,h,\gamma,t)$ for the $\Sf$ geometry (see 
		Fig.~\ref{fig:defstrucflat}) for ${T=T_c}$, $d=4$, and for various wedge angles $\gamma$
		as a function of $\Ltip/h=(L/h)-1$.
		$f_\perp^\Sf$ is rescaled by the factor $\big[h^d/\Delta_{+,+}\big]$, such that the 
		plotted quantity is	dimensionless and positive.
		All data corresponding to $\gamma=90\degree$ are multiplied by $10^{-1}$, the ones
		corresponding to $\gamma=37\degree$ are multiplied by $10^{-2}$.
		The gray lines correspond to the effective planar wall behavior [Eq.~\eqref{eq:crit-force-sf-eff}]
		with a decay $\propto \Leff^{-d}$, which is reached for $\Ltip/h\gg1$, i.e., at far distances;
		$\Leff/h=\Ltip/h+1-\hefftilde$.
		At small distances the actual data exhibit a crossover to a nearby regime 
		with a variation $\propto \Ltip^{-(d-1)}$ for $\Ltip/h\ll1$.
		(For $\gamma=37\degree$	this crossover occurs only for very small values of $\Ltip/h$ below 
		which the numerical	analysis is hardly feasible.)
		The Derjaguin-like approximation for the nearby regime [Eq.~\eqref{eq:derjaguin-2}], 
		which does not account for different values of $\gamma$, corresponds to a force behavior
		$f_\perp^{(Der)}\times\big[h^d/\Delta_{+,+}\big]=(\Ltip/h)^{-(d-1)}$.
		This behavior is exhibited qualitatively by the full numerical data.
		However, the amplitude depends on $\gamma$, which is described by introducing an additional factor
		$S(\gamma)$.
		From fits to the full numerical data we find for these factors the values $S(\gamma=139\degree)=1.0$,
		$S(\gamma=90\degree)=1.4$, and $S(\gamma=37\degree)=3.5$; these correspond to the 
		solid lines in the figure.
		For large wedge angles the (local) Derjaguin-like approximation agrees also quantitatively with 
		the full numerical data, whereas for small values of $\gamma$ the non-local character 
		of the critical Casimir force is stronger.
		}
	\label{fig:casimir-nearby}
\end{figure}

The full numerically obtained data for the normal critical Casimir force
are in qualitative agreement with the Derjaguin-like approximation (see Fig.~\ref{fig:casimir-nearby}).
However, this approximation [Eq.~\eqref{eq:derjaguin-2b}] does not capture the variation of
the amplitude of the force as a function of $\gamma$, which we find from the full numerical results
[Fig.~\ref{fig:casimir-nearby}].
%
%
%
%
\par
Finally, the analysis reveals that for the normal critical Casimir force between a corrugated substrate
and a flat substrate there is a {crossover} from a distant effective planar wall regime 
($f_\perp^\Sf\propto \Leff^{-d}$ for the case ${T=T_c}$)
to a nearby regime ($f_\perp^\Sf\propto \Ltip^{-(d-1)}$) [see Fig.~\ref{fig:casimir-nearby}].
We find that the distance at which this crossover occurs is varying with the wedge angles $\gamma$:
for small $\gamma$, this crossover occurs only at very small values of $\Ltip/h\ll1$, 
whereas for large $\gamma$ even for separations $\Ltip/h\simeq1$ the critical Casimir force varies
according its nearby behavior [Fig.~\ref{fig:casimir-nearby}].
%
\subsection{Lateral critical Casimir force for two identically structured substrates \label{sec:lateral}}
\begin{figure}

	\includegraphics{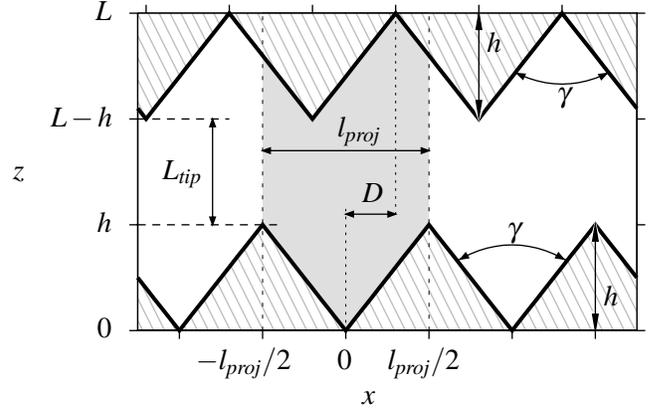}
	\caption{Confinement consisting of two identically geometrically \emph{s}tructured substrates
	(denoted as $\Ss$ geometry).
	Both structures are periodic in $x$ direction, but they are shifted by $D$ with respect
	to each other.
	}
	\label{fig:defstrucstruc}
\end{figure}

%
If two geometrically structured surfaces oppose each other, in addition to the normal critical Casimir forces
there are also lateral critical Casimir forces acting on the substrates.
Lateral forces are a particularly sensitive probe of the influence of roughness because there is no 
underlying planar wall contribution.
We restrict our study to the case in which both substrates are identically structured but laterally shifted 
with respect to each other as shown in Fig.~\ref{fig:defstrucstruc}.
%
%
The normal critical Casimir force for this geometry depends on the lateral shift $D$, 
but its distant behavior is similar to the effective planar wall limit discussed in Subsec.~\ref{sec:normal} 
[Eqs.~\eqref{eq:force-sf-eff} and \eqref{eq:crit-force-sf-eff}], but here with an effective planar 
wall distance $\Leff=L-2\heff$.
\par
Therefore, in this subsection we focus our attention on lateral critical Casimir forces.
The lateral force [Eq.~\eqref{eq:flateraldef}] for this $\Ss$ geometry exhibits the scaling form
\begin{align}
	\label{eq:crit-force-ss-1}
	f_{\parallel}^\Ss(D,L,h,\gamma,t)	
	&=\frac{d-1}{L^{d}}  \widetilde{f}_{\parallel,\pm}^{\,\,\Ss}(D_\pm,y_\pm,h_\pm,\gamma)\\
	&\xrightarrow{t\to0}\frac{d-1}{L^{d}}
				\Delta_{+,+}\widetilde{\Delta}_{\parallel}^\Ss(\tfrac{L}{h},\tfrac{D}{h},\gamma),\nonumber
\end{align}
where $\widetilde{f}_{\parallel,\pm}^{\,\,\Ss}$ and $\widetilde{\Delta}_\parallel^\Ss$ are the scaling
functions and the generalized Casimir amplitude for the lateral force, $D_\pm=D/\xi_\pm$ is the lateral
shift scaled by the correlation length, and $y_\pm=t(L/\xi_0^\pm)^{1/\nu}$ is the scaled distance variable.
For ${T=T_c}$ the dimensionless quantity ${f_{\parallel}^\Ss\times h^d}$ depends only on $L/h$, $D/h$, 
and $\gamma$.
Due to symmetry reasons it is sufficient to consider the lateral shift range 
$0\le\delta\le1/2$, where $\delta=D/\lproj$.
\par
As will be shown below, the lateral force decays exponentially 
(i.e., stronger than $L^{-d}$) with the wall distance $L$.
This implies that
\begin{equation}
	\widetilde{f}_{\parallel,\pm}^{\,\,\Ss}\,,\,\widetilde{\Delta}_\parallel^\Ss
		\xrightarrow{L\gg h,\lproj}0.
\end{equation}
The exponential decay of the lateral force as a function of the wall separation $L$ for values of $L$ that are much larger than
the corrugation height $h$ is expected to be present for any geometry of the corrugation.
However, as a function of the \emph{lateral} coordinate $D$, the behavior of the lateral force in the regime where the two plates approach
each other depends strongly on the details of the geometry.
One can even think of geometries for which the lateral force as a function of $D$ decays algebraically for $D\ll\lproj$.
Consider the 'extreme' corrugation geometry of peaks appearing periodically, like a comb geometry with all teeth but every fifth tooth removed.
If two such structured substrates approach each other such that $L<2h$, i.e., $\Ltip<0$, and $0<D\ll\lproj$, the main contribution to
the lateral force will be due to the critical Casimir force acting on two vertical walls on a horizontal support.
This force is the same as the 'normal' critical Casimir force between two planar walls, and it decays algebraically as a function of the lateral
distance as long as $D\ll\xi$.
For $D\simeq\lproj/2$ the influence of the neighboring corrugation peak becomes strong and the behavior of the lateral force is different
from the planar wall behavior.
This consideration points out that the lateral force can be viewed to be similar to a normal force but acting on tilted surfaces.
\subsubsection{Scaling functions within mean field theory}
%
We calculate numerically the scaling functions in Eq.~\eqref{eq:crit-force-ss-1} within MFT ($d=4$).
We find that for a fixed distance $L$ the free energy of the critical 
fluid in the confined $\Ss$ geometry takes its minimal value at $\delta=0$, i.e.,
the configuration of opposing tips of the corrugations is the preferred one.
The configuration $\delta=1/2$, where ridges are opposing wedges, corresponds to 
the maximal value of the free energy.
Therefore, the lateral critical Casimir force is negative (i.e., pointing in negative $x$ direction)
in the range $0<\delta<1/2$ and zero for $\delta=0,1/2$.
The lateral force is symmetric around the point $(\delta=1/2,f_{\parallel}^\Ss=0)$ and positive
in the range $1/2<\delta<1$.
For the $(+,+)$ case discussed here the 'in-phase' configuration $\delta=0$ is always the stable one. 
However, if the substrates are reconfigured with opposing surface fields ($(+,-)$, see, e.g., Refs.~~\onlinecite{KRECH94,KRECH97}),
critical Casimir forces change from being attractive to being repulsive, and we expect the preferred configuration of the two 
substrates in the $\Ss$ geometry to be the one in which the crest of an edge of one 
substrate opposes the center of a wedge of the other substrate, i.e., an 'out-of-phase' configuration with $\delta=1/2$, 
in contrast to the $(+,+)$ case.
\par
\begin{figure}

	\includegraphics{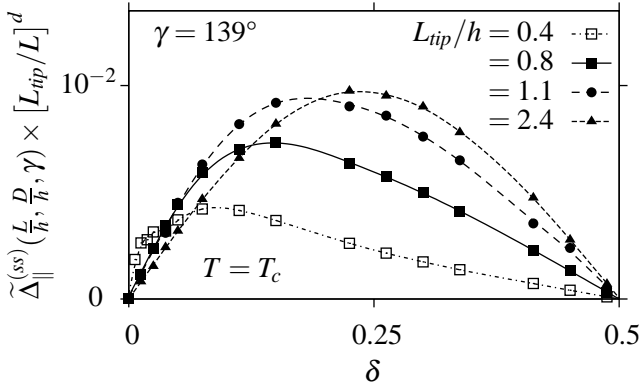}
	\caption{%
			The generalized Casimir amplitude 
			$\widetilde{\Delta}_\parallel^\Ss$
			of the lateral critical Casimir force [Eq.~\eqref{eq:crit-force-ss-1}] 
			for the $\Ss$ geometry (see Fig.~\ref{fig:defstrucstruc})
			within MFT ($d=4$)
			as a
			function of the lateral shift $\delta=D/\lproj$.
			The amplitudes are
			multiplied by $\big[\Ltip/L\big]^d$ in order to facilitate a quantitative comparison with 
			the normal critical Casimir force between two planar walls at distance $\Ltip$.
			This shows that the lateral critical Casimir force is about a factor of $100$ smaller than
			its normal counterpart.
			Here the wedge angle is $\gamma=139\degree$ which corresponds to $\lproj=5.33h$.
			For $\Ltip\lesssim \lproj/2$ the shape of the scaling function for the lateral force 
			is asymmetric around $\delta=0.25$,
			whereas for larger distances the shape becomes sinusoidal.
		}
	\label{fig:casimir-lateral}
\end{figure}

The shape of the lateral critical Casimir force as a function of $\delta$ changes upon varying the 
distance between the walls as shown in Fig.~\ref{fig:casimir-lateral}.
%
%
Whereas for $\Ltip\lesssim \lproj/2$ the shape of the generalized Casimir amplitude is asymmetric around 
$\delta=1/4$, it becomes sinusoidal for $L\gtrsim \lproj$.
This asymmetry for small wall distances reflects the details of the geometry of the corrugation, whereas
the symmetric sinusoidal shape depends only on the major parameters of the corrugation, 
namely its periodicity and its height.
\par
\begin{figure}

	\includegraphics{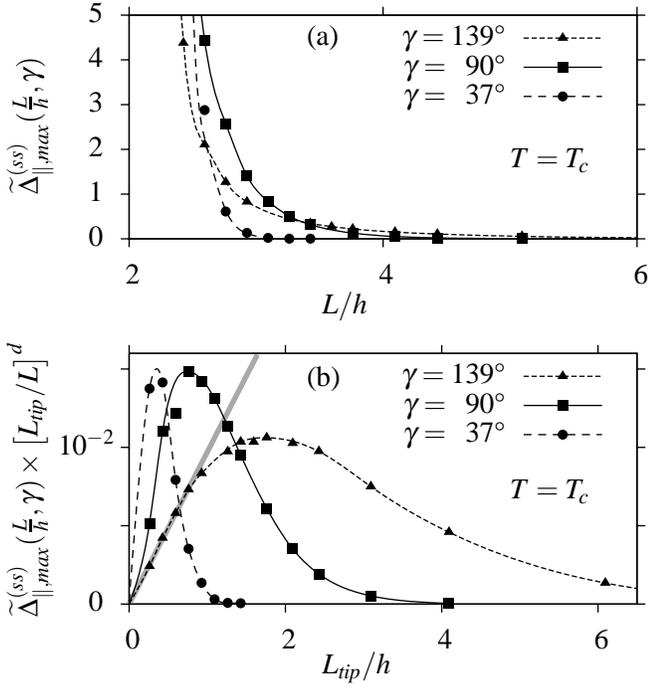}
	\caption{%
		(a)
		Amplitude $\widetilde{\Delta}_{\parallel,\Max}^\Ss$ (i.e., maximal absolute 
		value with respect to the lateral shift $D$) of the lateral critical Casimir force 
		scaling function at criticality for the $\Ss$ geometry (see Fig.~\ref{fig:defstrucstruc}) 
		within MFT ($d=4$) as a function of $L/h$ for different wedge angles $\gamma$.
		The values of $\widetilde{\Delta}_{\parallel,\Max}^\Ss$ indicate the strength of the lateral
		force compared to the normal critical Casimir force between two planar walls at distance $L$
		(see Eq.~\eqref{eq:crit-force-ss-1} and Fig.~\ref{fig:defstrucstruc}); 
		their magnitudes may be similar.
		(b)	Generalized Casimir amplitude $\widetilde{\Delta}_{\parallel,\Max}^\Ss$
		multiplied by $[\Ltip/L]^d$ as a function of $\Ltip/h=L/h-2$.
		This product is the ratio of the maximal lateral force at distance $L$ and the normal
		force between planar walls at distance $\Ltip$ (see Eq.~\eqref{eq:crit-force-ss-1}).
		The fact that these curves vanish linearly for $\Ltip\to0$ (see the shaded straight line)
		indicates that $\widetilde{\Delta}_\parallel^\Ss\propto \Ltip^{-(d-1)}$ for $\Ltip\to0$,
		analogous to the normal force [Eq.~\eqref{eq:derjaguin-2}].
		This implies that in (a) $\widetilde{\Delta}_{\parallel,\Max}^\Ss(L/h\to2)\propto(L/h-2)^{-(d-1)}$.
		For small values of $\Ltip/h$ the lateral critical Casimir force is stronger for
		corrugations with a short corrugation wavelength $\lproj$, i.e., at small angles $\gamma$.
		On the other hand, the amplitudes of the lateral force for these short wavelength corrugations
		decay faster for $\Ltip/h\gg1$.
		The maximum of the lateral critical Casimir force at distance $L$ is of the order of 
		$\mathcal{O}(0.01)$ compared to the normal critical Casimir force between two planar walls at 
		distance $\Ltip$.
		The curves both in (a) and (b) decay exponentially for $L\to\infty$ (see Fig.~\ref{fig:lateral-amp}).
		}
	\label{fig:lateral-amp-scaled}
\end{figure}

We define the maximum of the lateral critical Casimir force 
$f_{\parallel,\Max}^\Ss(L,h,\gamma,t)$ to be its maximal absolute value in the range $0\le\delta\le1/2$.
The corresponding maxima of the universal scaling functions of the lateral force are
$\widetilde{f}_{\parallel,\Max}^{\,\,\Ss\pm}(y_\pm,h_\pm,\gamma)$ and
$\widetilde{\Delta}_{\parallel,\Max}^\Ss(L/h,\gamma)$.
Typical values of $\widetilde{\Delta}_{\parallel,\Max}^\Ss$ range from $0$ to $\mathcal{O}(100)$, 
depending on the distance between the two walls (see Fig.~\ref{fig:lateral-amp-scaled}(a)).
This means that the lateral critical Casimir force can be much stronger than the normal critical Casimir
force between two planar walls at distance $L$.
However, $L$ is not necessarily the relevant distance to compare with (see Fig.~\ref{fig:defstrucstruc}).
Compared with the magnitude of the normal critical Casimir force between two planar walls at 
distance $\Leff$, the maximum of the lateral force can be of similar strength.
The magnitude of the lateral force is of the order of $\mathcal{O}(0.01)$ compared to the normal critical 
Casimir
force between two planar plates at distance $\Ltip$ as long as the corrugated walls are not too far 
away from each other (see Fig.~\ref{fig:lateral-amp-scaled}(b)).
For $\Ltip\gg \lproj$ the lateral Casimir force vanishes relative to the normal force.
Concerning the behavior for different corrugation patterns we find that for $\Ltip\lesssim h$ 
the maxima of the lateral critical Casimir force are larger for small values of $\gamma$
(Fig.~\ref{fig:lateral-amp-scaled}(b)).
On the other hand, the lateral force for small values of $\gamma$ vanishes 
faster upon increasing the wall distance.
%
%
%
%
\par
\begin{figure}

	\includegraphics{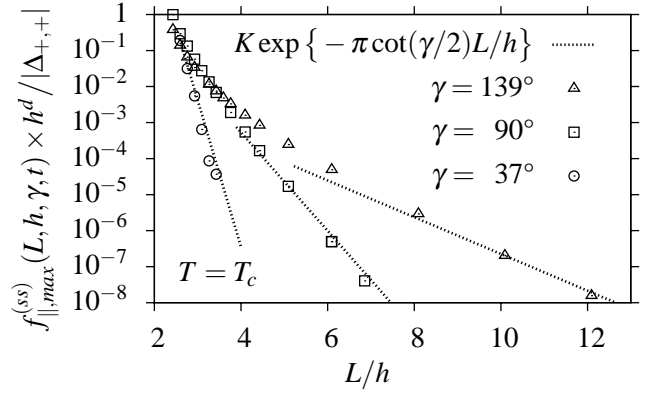}
	\caption{%
	    Maximal absolute value of the lateral critical Casimir force [Eq.~\eqref{eq:crit-force-ss-1}]
		for the $\Ss$ geometry [Fig.~\ref{fig:defstrucstruc}]
		at ${T=T_c}$ rescaled by $\big[h^d/|\Delta_{+,+}|\big]$ such that the plotted quantity is
		dimensionless.
		The lateral force decays exponentially with $L/h$ for all
		wedge angles $\gamma$, but with different characteristic decay constants.
		The exponential decay for $L\gg h,\lproj$ agrees reasonably well with the predictions
		from the superposition approximation [Eq.~\eqref{eq:caslat-final-2}] (dotted lines).
		From fits to the curves we find for the dimensionless prefactor $K$ in 
		Eq.~\eqref{eq:caslat-final-2} the values 
		$K(\gamma=139\degree)=2.8\times10^{-2}$,
		$K(\gamma=90\degree)=1.5\times10^{2}$,
		and $K(\gamma=37\degree)=7.2\times10^{9}$.
		}
	\label{fig:lateral-amp}
\end{figure}

From Fig.~\ref{fig:lateral-amp-scaled}(b) we infer that the
nearby behavior (for $\Ltip/h\to0$) of the lateral critical Casimir force maximum 
is dominated by a behavior $\propto \Ltip^{-(d-1)}$, 
similar to the normal force [Eq.~\eqref{eq:derjaguin-2}].
However, there is a pronounced difference to the normal force concerning the distant behavior.
Whereas the normal force approaches an effective planar wall limit with an algebraic decay
$\propto \Leff^{-d}$ at ${T=T_c}$ [Fig.~\ref{fig:casimir-zero}], the lateral
force decays {exponentially} for $T\neq T_c$ as well as for ${T=T_c}$ [Fig.~\ref{fig:lateral-amp}].
%
%
The exponential decay of the lateral critical Casimir force can be explained in terms
of a superposition approximation for large distances between the walls.
If one uses the approximative forms of the order parameter profile at a single 
corrugated substrate obtained in Subsec.~\ref{sec:undulation} and calculates the stress tensor
components one arrives at (see Appendix~\ref{sec:appendix-2}) 
\begin{multline}	
	\label{eq:caslat-final}
	f_{\parallel}^\Ss(D,L\gg h,h,\gamma,t=0)\simeq
	\\
				-K\frac{|\Delta_{+,+}|}{h^d}\;
				e^{-\frac{2\pi L}{\lproj} }
				\sin\left(\tfrac{2\pi D}{\lproj} \right),
\end{multline}
where $K$ is dimensionless, positive, and depends only on $\gamma$.
This form captures the sinusoidal distant behavior $\propto-\sin(2\pi\delta)$ 
of the lateral critical Casimir force for fixed $L$ as well as the free energetic
preference of the opposing tips configuration.
Equation \eqref{eq:caslat-final} implies the exponential decay of the maximum of the lateral force:
\begin{multline}
	\label{eq:caslat-final-2}
	f_{\parallel,\Max}^\Ss(L\gg h,h,\gamma,t=0)\simeq
	\\ 
					K\frac{|\Delta_{+,+}|}{h^d}\;
					\exp\left\{-\pi\cot\left(\frac{\gamma}{2}\right)\frac{L}{h}\right\}.
\end{multline}
Although the 
prefactor $K$ in the appropriate expression in Eq.~\eqref{eq:caslat-final} cannot be given analytically,
the decay constants found using this superposition approximation agree well with the numerical data 
for $L/h\gg1$ (see Fig.~\ref{fig:lateral-amp}).
The exponential decay constant in Eq.~\eqref{eq:caslat-final} corresponds to the 'reciprocal
lattice vector' of the periodic array of wedges and ridges.
It is worth mentioning that the lateral electrodynamic Casimir force due to corrugations of the walls 
decays exponentially for large wall distances with the reciprocal corrugation lattice vector as decay 
constant \cite{EMIG03,EMIG05}.
\par
\begin{figure}

	\includegraphics{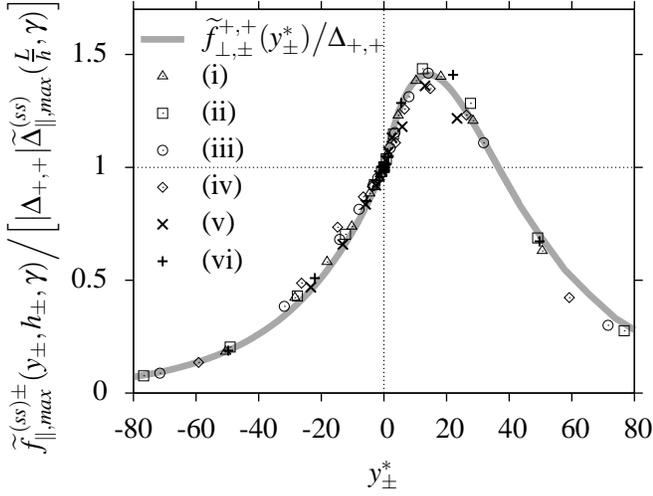}
	\caption{
		Maximum (with respect to $D$) of the scaling function of the lateral critical Casimir force
		for the $\Ss$ geometry (see Fig.~\ref{fig:defstrucstruc}) in units of its value at ${T=T_c}$ 
		[Eqs.~\eqref{eq:crit-force-ss-1} and \eqref{eq:tempdependence}]	as a function of 
		$y_\pm^*=t(L^*/\xi_0^\pm)^{1/\nu}$ (note that within MFT $y_+^*=\tau (L^*)^2$, and 
		$y_-^*=2\tau (L^*)^2$).
		$L^*$ is related to the tip-tip distance $\Ltip$ 
		[Fig.~\ref{fig:defstrucstruc}]
		as $L^*=W(\Ltip/h,\gamma)\;\Ltip$ [Eq.~\eqref{eq:L-star}],
		where $W$ is a dimensionless scaling factor to be determined (see below and the main text).
		The gray curve corresponds to the scaling function for the \emph{normal} force between two 
		\emph{planar} walls at distance $L^*$.
		The data points correspond to the substrate configurations
		$\gamma=90\degree$ and 
		(i) $\Ltip/h=1/2$,
		(ii) $\Ltip/h=1$,
		(iii) $\Ltip/h=2$,
		(iv) $\Ltip/h=3$,
		and to the configurations
		(v) $\gamma=37\degree$ and $\Ltip/h=1$, as well as
		(vi) $\gamma=139\degree$ and $\Ltip/h=4/3$.
		The numerically determined scaling factors $W$ for these cases are approximately 
		(i) $1.42$,
		(ii) $1.17$,
		(iii) $0.94$,
		(iv) $0.86$,
		(v) $0.81$, and
		(vi) $1.17$
		with an error of ca. $\pm0.05$ for all cases.
		With these rescaling factors all data fall onto the gray curve.
	}
	\label{fig:amplitudes-scaling}
\end{figure}

As a function of temperature the maximum strength of the lateral critical Casimir force is found numerically to  behave 
approximately as
\begin{equation}
	\label{eq:tempdependence}
	\widetilde{f}_{\parallel,\Max}^{\,\,\Ss\pm}(y_\pm,h_\pm,\gamma)
	\simeq
	\widetilde{\Delta}_{\parallel,\Max}^\Ss\left(\tfrac{L}{h},\gamma\right)
	\left|{\widetilde{f}_{\perp,\pm}^{+,+}(y_\pm^*)}\right|,
\end{equation}
where $\widetilde{f}_{\perp,\pm}^{+,+}(y_\pm)=\widetilde{f}_{\perp,\pm}^{\,\,\Sf}(y_\pm,h_\pm=0,\gamma=\pi)$ 
(see Eq.~\eqref{eq:crit-force-sf-1}) is the planar wall {normal} force scaling function
which is evaluated at $y_\pm=y_\pm^*=t (L^*/\xi_0^\pm)^{1/\nu}$.
This is shown in Fig.~\ref{fig:amplitudes-scaling}.
From the numerical data we infer that the characteristic distance $L^*$
is equal to the tip-tip distance $\Ltip$ apart from a dimensionless scaling factor $W$ which
depends on the ratio $\Ltip/h$ and the angle $\gamma$, i.e.,
\begin{equation}
	\label{eq:L-star}
	L^*=W\left(\tfrac{\Ltip}{h},\gamma\right)\;\Ltip.
\end{equation}
The numerical data for the examples presented in Fig.~\ref{fig:amplitudes-scaling} imply that
the dependence of $W$ on $\Ltip/h$ and $\gamma$ cannot be further reduced to a dependence on
$\Ltip/\lproj$ only.
We find typical values of $W$ to be in the range $0.8$ to $1.5$ for the examples shown in
Fig.~\ref{fig:amplitudes-scaling}.
However, it is rather interesting that the temperature dependence for the \emph{lateral} critical 
Casimir force between {corrugated} substrates can be related to the \emph{normal} critical Casimir 
force scaling function for \emph{flat} substrates in a simple way.
We interpret this aspect as a consequence of the lateral force to be basically 
a sort of 'normal force' between several tilted walls, generated by the same physical mechanisms.
%

%
\section{Summary \label{sec:summary}}
%
%
We have investigated the universal properties of strong 
critical adsorption of fluids on geometrically structured substrates
which are modeled by an array of wedges and ridges characterized
by the corrugation height $h$ and the wedge angle $\gamma$ [Fig.~\ref{fig:definitions}].
These universal properties  correspond to the physical quantities of a system whenever all relevant 
length scales are larger than their characteristic molecular length scales.
Moreover, we have studied the singular contributions to the normal and lateral critical Casimir
forces acting on geometrically structured substrates which confine a fluid close to criticality.
\subsection{Critical adsorption}
\setcounter{summary}{0}
Concerning the study of critical adsorption at a single corrugated substrate, in
Sec.~\ref{sec:criticaladsorption} we have obtained the following main results based on general scaling 
arguments and explicit mean field calculation for scaling functions.
\sumnum
Close to the bulk critical point at ${T=T_c}$ the order parameter profile can be described in 
terms of universal scaling functions depending on length variables scaled by the 
correlation length $\xi_\pm$ above ($+$) or below ($-$) $T_c$ and the wedge angle $\gamma$ 
[Eq.~\eqref{eq:scaling}].
Within the scaling limit the order parameter profile diverges close to the walls according to a power law [Eq.~\eqref{eq:shortdistance}].
For $T\neq T_c$ it decays exponentially into the bulk [Eq.~\eqref{eq:scaling-decay}].
At criticality the order parameter profile reduces to a pure power law in the direction $z$
perpendicular to the mean substrate surface,
multiplied by a universal amplitude function [Eq.~\eqref{eq:scaling-zero}] which depends 
on $x/h$, $z/h$, and $\gamma$, where
$x$ is the direction of periodicity of the corrugation.
\sumnum
The distant behavior far from the wall is governed by an effective planar wall behavior, i.e., the 
order parameter profile resembles the one generated by a planar wall located at the effective 
height $z=\heff$ [Eq.~\eqref{eq:heff-decay}] with 
$\heff=h\;\hefftilde$ [Eq.~\eqref{eq:def-hefftilde}].
\sumnum
We have calculated numerically the scaling functions within mean field theory 
for $T< T_c$, ${T=T_c}$, and ${T>T_c}$  
[Figs.~\ref{fig:orderparamprofiles}, \ref{fig:contourcompare}, \ref{fig:contourzero}, 
\ref{fig:cut}, and \ref{fig:heff}].
Approximately, the effective height is given by 
$\hefftilde\simeq1-\gamma/2\pi$ [inset of Fig.~\ref{fig:heff}].
\sumnum
In order to compare the proliferation of the surface structure into the bulk for the cases ${T<T_c}$, ${T=T_c}$, 
and ${T>T_c}$, we have introduced the width $\Delta s(s)$
of the undulation of the order parameter contour lines with mean position $z=hs$ [Fig.~\ref{fig:spatvar}].
Within mean field theory we have calculated $\Delta s(s)$ using a superposition approximation 
serving as a distant wall approximation.
We have found that the surface structure imprinted onto the liquid decays exponentially into the bulk 
for all temperatures, including $T_c$ [Eq.~\eqref{eq:delta}].
This is confirmed by the numerical data.
The analytically determined corresponding exponential decay constants 
[Eqs.~\eqref{eq:sigma0} and \eqref{eq:sigma}]
compare well with the numerical data [Fig.~\ref{fig:spatvarcompare}].
Interestingly, the lateral structure of the order parameter decays into the bulk 
fastest at ${T=T_c}$, in contrast to the order parameter profile itself.
\sumnum
We have introduced the excess adsorption 
as the integral of the difference of the order parameter and its bulk
value over a unit cell of the periodic structure of the system.
For a binary liquid mixture this is a measure of the total enrichment of a species
at the substrate.
Close to $T_c$ the singular part of the excess adsorption $\Gamma$ exhibits scaling [Eq.~\eqref{eq:excess-struc}]
with the corresponding universal scaling function $\widetilde{\Gamma}_\pm$ depending only on $h/\xi_\pm$ 
and $\gamma$.
In order to compare the excess adsorption at a corrugated substrate with the one at a flat substrate, 
we have defined a suitably defined reduced 
relative excess adsorption [Eq.~\eqref{eq:upsilon-definition}].
\sumnum
We have calculated the universal scaling functions for the excess adsorption within mean field theory 
and have compared them with the one for planar walls [Figs.~\ref{fig:side} and \ref{fig:proj}].
We have found that the amount of adsorbed matter is less than the one at a planar wall of the same 
actual substrate surface. 
This effect is increasing for decreasing angles $\gamma$.
Compared to a planar substrate with the same \emph{projected} area the excess adsorption on a 
corrugated substrate is larger.
For $\gamma\to0$ the ratio of these latter two adsorptions even diverges, and this effect 
is enhanced upon departing from criticality.
We have defined a suitable ratio of the excess adsorptions above and below $T_c$ [Eq.~\eqref{eq:R-definition}]
and have found that it may be smaller than the corresponding planar wall ratio, 
depending on the wedge angle $\gamma$.
For $\gamma\to0$ the difference between these two ratios is of the order of $10\%$ [Fig.~\ref{fig:ratio}].
\subsection{Critical Casimir forces}
\setcounter{summary}{0}
In Sec.~\ref{sec:casimir} we have investigated critical Casimir forces between geometrically structured 
substrates by considering two arrays of wedges and ridges separated
by a distance $L$ in $z$ direction and periodic in $x$ direction [Figs.~\ref{fig:defstrucflat} and 
\ref{fig:defstrucstruc}].
We have focused on identical chemical boundary conditions on both substrate surfaces.
In the following our main findings are summarized.
\sumnum
Close to criticality the normal and the lateral critical Casimir forces obey scaling behaviors 
[Eqs.~\eqref{eq:crit-force-sf-1} and \eqref{eq:crit-force-ss-1}].
In order to determine the corresponding universal scaling 
functions for ${T<T_c}$, ${T=T_c}$, and ${T>T_c}$, we have applied
the stress tensor method.
It turns out that the improvement term to the stress tensor is
vanishing for periodic geometries.
The order parameter profiles and the universal scaling functions for the critical Casimir forces
have been calculated numerically within mean field theory for the following two basic configurations.
\subsubsection{{Normal critical Casimir force}}
\setcounter{summary}{0}
\subsumnum 
First, we have considered the configuration of a geometrically structured substrate opposing a planar wall
[Fig.~\ref{fig:defstrucflat}].
In this case there is a normal critical Casimir force acting on the substrates.
We have found a crossover between two limiting regimes. 
\subsumnum
The distant wall regime $L/h\gg1$ is governed by an effective planar wall behavior, for which the scaling 
function of the normal force attains the form of the one between two planar walls separated by the
distance $L-\heff$,
where $\heff$ turns out to be the same as for a single corrugated substrate
[Figs.~\ref{fig:casimir-normal-distant} and \ref{fig:casimir-zero}].
\subsumnum
On the other hand, the nearby regime exhibits an algebraic 
behavior $\propto (\Ltip)^{-(d-1)}$, where $\Ltip=L-h$ is the tip-wall distance
[Fig.~\ref{fig:casimir-nearby}].
This behavior is different from the one between two planar walls, which would be $\propto H^{-d}$, 
if $H$ is the wall distance.
Using a Derjaguin-like approximation, we have explained this behavior 
qualitatively [Eqs.~\eqref{eq:derjaguin-1} and
\eqref{eq:derjaguin-2}].
\subsubsection{{Lateral critical Casimir force}}
\setcounter{summary}{0}
\subsumnum
Second, we have considered the case in which two identically corrugated substrates are opposing 
each other, but possibly shifted laterally with respect to each other by a distance $D$ along the $x$ 
direction [Fig.~\ref{fig:defstrucstruc}].
The universal scaling function of the resulting lateral critical Casimir force acting on the substrates
depends on $D/\xi_\pm$, $L/\xi_\pm$, $h/\xi_\pm$, and $\gamma$ for $T\neq T_c$, and on
the scaled variables $D/h$, $L/h$, and  $\gamma$ for ${T=T_c}$, respectively.
\subsumnum
We have found that the configuration of opposing tips is the preferred one and the configuration of a wedge 
opposing a tip corresponds to an unstable configuration.
\subsumnum
The shape of the scaling function of the lateral force in between those two configurations depends 
on the relative shift $D$ between the walls
[Fig.~\ref{fig:casimir-lateral}].
For wall distances which are small compared to the projected width of a wedge or the height of the wedges
the lateral force reflects the detailed geometry of the system.
On the other hand, for distances larger than $h$ and the projected width of a wedge
the shape becomes sinusoidal [Fig.~\ref{fig:casimir-lateral}].
\subsumnum
The amplitude of the lateral critical Casimir force depends on $L$ but there is also a strong
dependence on the wedge angle $\gamma$ [Fig.~\ref{fig:lateral-amp-scaled}].
Within a suitable comparison scheme the amplitude 
of the lateral critical Casimir force turns out to be of the order of $1\%$ 
of the one for normal critical Casimir forces between two planar walls at distance $L-2h$.
There is evidence that in the nearby regime the lateral force amplitude 
scales as $(L-2h)^{-(d-1)}$
[Fig.~\ref{fig:lateral-amp-scaled}(b)].
The distant behavior is characterized by an exponential decay of the lateral force amplitude into the bulk
[Fig.~\ref{fig:lateral-amp}].
\subsumnum
Using a superposition approximation, we have been able to explain the sinusoidal shape of the lateral force, 
the free energy behavior, and the exponential decay of the force amplitude for large distances 
[Eq.~\eqref{eq:caslat-final}].
The approximatively calculated decay constants compare well with the full numerical 
data [Fig.~\ref{fig:lateral-amp}].
\subsumnum
We have found that the temperature dependence of the maximal strength of the \emph{lateral} critical Casimir force can be expressed in terms 
of the temperature dependence of the \emph{normal} critical Casimir force between planar walls [Eq.~\eqref{eq:tempdependence} and 
Fig.~\ref{fig:amplitudes-scaling}].
This can be interpreted in the sense that the lateral critical Casimir force between corrugated substrates 
can be viewed to be a normal critical Casimir force between tilted walls, and thus being generated
by the same physical mechanisms as the normal critical Casimir force.
\subsubsection{Comparison with the quantum electrodynamic Casimir effect}
\setcounter{summary}{0}
Surface corrugations lead to pronounced effects on the \emph{quantum electrodynamic} Casimir force, too
\cite{Emig:2001,Zwol:2008,EMIG03,Chen:2002,EMIG05,Rodrigues:2006,Dalvit:2008,MIRI07,
EMIG07,Miri:2008,Rodrigues:2008,Chan:2008,Lambrecht:2008}.
\subsumnum
Walls, which are geometrically structured, lead to a behavior of the normal quantum electrodynamic Casimir
effect, which is different from the one for planar walls \cite{Emig:2001,EMIG03,Zwol:2008,Chan:2008,Lambrecht:2008}.
Generally, the normal quantum electrodynamic Casimir force is enhanced compared with the one between
two planar walls separated by the same mean distance, and weakened compared with the one between two 
planar walls at the same minimal, i.e., tip-to-tip distance.
This is similar to the behavior of the normal critical Casimir force 
[Fig.~\ref{fig:casimir-normal-distant}(a)], which we also found to be enhanced compared with the one acting 
between planar walls separated by the same mean distance, and weakened compared with the one acting on
planar walls at distance $\Ltip$ [Fig.~\ref{fig:defstrucflat}].
\subsumnum
The effective planar wall limit of the normal critical Casimir force 
[Fig.~\ref{fig:casimir-normal-distant}(b)] with the effective corrugation height $\heff=h\,\hefftilde$
given by the expression $\hefftilde$ for semi-infinite systems [Eq.~\eqref{eq:heff-result}] is 
consistent with the results obtained for the normal quantum electrodynamic Casimir force 
between a periodically structured and a planar wall \cite{EMIG03}.
However, most corresponding studies for the quantum electrodynamic Casimir effect consider sinusoidal
or rectangular shaped corrugation profiles, which are different from the wedge-like shapes
which we have discussed upon extending previous studies of this type of geometry \cite{HANKE99,PALAGYI04}.
Therefore, a comparison of the nearby regimes of the two effects is difficult.
%
\subsumnum
The dependence of the amplitude of the lateral critical Casimir force for the $\Ss$ geometry on the wall
distance [Fig.~\ref{fig:casimir-lateral}], with a geometry dependent form in the nearby regime and a
sinusoidal form in the distant regime, is similar to the one for the lateral quantum electrodynamic Casimir 
force \cite{EMIG05}.
\subsumnum
Both the lateral critical Casimir force for geometrically structured substrates 
[Fig.~\ref{fig:casimir-lateral}] and the lateral quantum electrodynamic Casimir force are directed such that 
the preferred configuration of the two corrugated substrates is the one of opposing tips
\cite{EMIG03,Chen:2002,EMIG05,Rodrigues:2006}.
However, critical Casimir forces change from being attractive to being repulsive, 
if the substrates are endowed with opposing surface fields ($(+,-)$, see, e.g., Refs.~~\onlinecite{KRECH97,
Fukuto:2005,Hertlein:2008,Vasilyev:2007}) instead of the $(+,+)$ case discussed here.
For corrugated substrates corresponding to the $(+,-)$ case we expect the preferred 
configuration of the two substrates in the $\Ss$ geometry to be the one in which the crest of an edge of one 
substrate opposes the center of a wedge of the other substrate, in contrast to the $(+,+)$ case.
\subsumnum
The exponential decay of the lateral critical Casimir force amplitude for large distances
(see Eq.~\eqref{eq:caslat-final-2}) is consistent with the behavior of the lateral quantum 
electrodynamic Casimir force \cite{EMIG03,Rodrigues:2006}, too.
Both decay with the 'reciprocal lattice vector' corresponding to the periodic lateral structure as the 
characteristic inverse decay length.
\subsumnum
The nearby regimes of the lateral critical Casimir force and the lateral quantum electrodynamic Casimir force are difficult to compare, because most of the corresponding studies of the lateral quantum 
electrodynamic force deal with sinusoidally shaped corrugations, whereas the present study of the
lateral critical Casimir force corresponds to wedge-like shapes.
\subsumnum
Actuation of nano-mechanical devices by the quantum electrodynamic Casimir effect may be possible using
geometrically structured objects \cite{MIRI07,EMIG07,Miri:2008,Rodrigues:2008}.
We expect that similar applications may be achieved using critical Casimir forces.
However, the fact that the critical Casimir force can be changed from being attractive to being repulsive
upon changing the chemical boundary conditions opens up additional possibilities.
\subsection{Discussion and outlook}
\begin{figure}

	\includegraphics{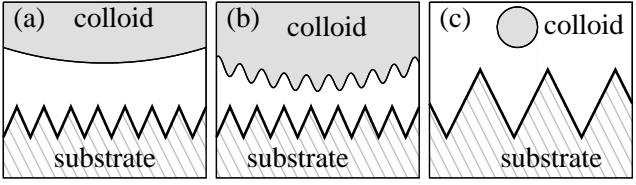}
	\caption{
		Possible experimental realizations for the measurement of critical Casimir forces
		for geometrically structured confinements (schematic).
		Commonly, in experiments colloids or spheres are used instead of two perfectly aligned
		planar walls (see, e.g., Refs.~~\onlinecite{Hertlein:2008,Chen:2002}). 
		Here the colloids are considered to be long rods.
		(a) A 'large' colloid opposing a geometrically structured substrate could provide access to the
		normal critical Casimir force for the $\Sf$ geometry [Fig.~\ref{fig:defstrucflat}].
		(b) A colloid with imprinted lateral sinusoidal corrugations opposes a geometrically structured
		substrate (a similar setup has been realized for the quantum electrodynamic Casimir force
		\cite{Chen:2002}).
		In such a geometry, lateral critical Casimir forces should arise similar to the ones
		arising in the $\Ss$ geometry [Fig.~\ref{fig:defstrucstruc}].
		(c) A 'small' colloid which is comparable in size with the corrugation height and the lateral
		extension of a single wedge could provide access to the normal or the lateral 
		critical Casimir force acting on a single 'tip' of the upper substrate in 
		Fig.~\ref{fig:defstrucstruc}.
		}
	\label{fig:experiments}
\end{figure}

\setcounter{summary}{0}
\sumnum
In addition to the critical Casimir forces due to fluctuations of the order parameter, there are
the omnipresent van der Waals dispersion forces acting as effective background forces on the substrates 
confining the fluids.
The total force is approximately \cite{Dantchev:2006,Dantchev:2007} the sum of these background forces 
and the critical Casimir force.
However, it has been shown theoretically \cite{Schlesener:2003} and 
experimentally (see, e.g., Ref.~~\onlinecite{Hertlein:2008}) that 
for classical fluids (corresponding to the presence of symmetry breaking surface fields)
the normal critical Casimir forces for $T\to T_c$ are much larger than these background forces 
and are directly accessible.
The same is expected to hold for the \emph{normal} critical Casimir force between corrugated substrates.
We have derived an estimate for the strength of the \emph{lateral} critical Casimir force
compared with the strength of the lateral dispersion background forces for such systems
(see Appendix~\ref{sec:vdw}).
Accordingly, for typical physical parameters [Subappendix~\ref{sec:num}] and upon approaching criticality
the lateral critical Casimir force is much larger than the corresponding lateral van der Waals force.
\sumnum
Lateral critical Casimir forces also occur if simple fluids \cite{SPRENGER06B} or liquid crystals 
in their nematic phase \cite{Karimi:2006} are confined by \emph{chemically} structured substrates.
Also in these systems the lateral force is sinusoidally shaped for large substrate distances
and depends on the details of the pattern structure for small substrate distances.
Similar to the lateral critical Casimir force for the $\Ss$ geometry (see Fig.~\ref{fig:lateral-amp})
also in these systems the lateral forces decay exponentially for increasing wall distance
with the 'reciprocal lattice vector' as characteristic length scale \cite{Karimi:2006}.
The strength of the lateral critical Casimir force due to the confinement of critical fluids 
by two substrates patterned with chemical stripes is of the order of the normal critical Casimir force
between these substrates whenever the wall distance is small \cite{SPRENGER06B}.
This ratio is larger than the corresponding one of the lateral critical Casimir force for the $\Ss$ 
geometry compared to
the normal critical Casimir force between two plates at distance $\Ltip$ (see
Figs.~\ref{fig:defstrucstruc} and \ref{fig:lateral-amp-scaled}(b)). However, if one
compares the lateral critical Casimir force with the normal one between two planar walls of 
distance $L$
(see Figs.~\ref{fig:defstrucstruc} and \ref{fig:lateral-amp-scaled}(a))
or $\Leff$, 
one finds that its strength can be of similar
order (for $\Leff$) or even larger (for $L$) compared with the corresponding planar wall normal force.
Therefore, lateral critical Casimir forces due to corrugations may be comparable in magnitude 
with lateral critical Casimir forces due to chemical structures.
\sumnum
An experimental realization for the measurement of normal and lateral critical Casimir forces
acting on geometrically structured substrates would provide access to the geometrical structure
of the corrugation of buried solid-liquid interfaces.
Recently, the direct measurement of the critical Casimir force between a colloid and a planar wall
confining a binary liquid mixture has been reported \cite{Hertlein:2008}.
A similar experiment, in which the flat substrate is replaced by a corrugated one,
with a typical corrugation size much smaller than the
extension of the colloid, could provide access to the normal critical Casimir force in the 
$\Sf$ geometry (see Figs.~\ref{fig:experiments}(a) and \ref{fig:defstrucflat}).
%
%
The lateral quantum electrodynamic Casimir force has been measured between a sinusoidally corrugated
metal sphere and a corrugated planar wall \cite{Chen:2002}. 
An experiment involving a geometrically structured colloid could provide direct access to the
lateral critical Casimir force for the $\Ss$ geometry
(see Figs.~\ref{fig:experiments}(b) and \ref{fig:defstrucstruc}).
Such setups  can play a role for moving ratchet-like parts in nano-machines. 
Using a smooth colloid which is small compared with the corrugation, one may, for example, measure the 
individual local
contributions to normal and lateral critical Casimir force [Fig.~\ref{fig:experiments}(c)].
%
%
\section*{Acknowledgment}
The authors thank an anonymous referee for raising the issue of lateral critical Casimir forces decaying algebraically
as a function of the lateral coordinate.
%
%
\appendix
\section{Undulation of the order parameter contour lines \label{sec:appendix-1}}
%
The behavior of the undulation $\Delta s(s)$ of the contour lines of the order
parameter profile close to a single corrugated substrate (see Subsec.~\ref{sec:undulation}) is studied
within MFT and a distant wall approximation scheme by introducing a superposition approximation for the
order parameter [Eqs.~\eqref{eq:superposition} and \eqref{eq:product}].
%
\subsection{Distant behavior of the order parameter profile for ${T=T_c}$ \label{sec:appendix-zero}}
%
Within MFT the order parameter profile ${m(x,z,h,\gamma,t=0)\equiv m_0(x,z)}$ at criticality solves the
differential equation [Eq.~\eqref{eq:diffeq}]
\begin{equation}
	\label{eq:m0-deq}
	\nabla^2 m_0(x,z)=(m_0(x,z))^3.
\end{equation}
Far from the substrate, i.e., for $z/h\gg1$, the order parameter profile contribution $m_0^{(1)}(x,z)$
due to the corrugation and the underlying planar wall contribution $m_0^{(0)}(z)$ 
[Eq.~\eqref{eq:superposition}] are related as
\begin{equation}
	m^{(1)}_0(x,z)\ll m^{(0)}_0(z).
\end{equation}
Thus, for the cubic power of $m_0$ one can neglect quadratic or higher order terms of $m_0^{(1)}$.
Within MFT the underlying effective planar wall contribution [Eqs.~\eqref{eq:heff-decay} 
and \eqref{eq:definition-heff}] is
\begin{equation}
	\label{eq:m00-heff}
	m_0^{(0)}(z)=c_+ (z-\heff)^{-1},
\end{equation}
where $c_+$ is the universal surface amplitude for the planar wall system [Eq.~\eqref{eq:definition-heff}]
with $c_+=\sqrt{2}$ within MFT \cite{FLOETER95}.
Together with Eq.~\eqref{eq:product} (i.e., the product ansatz $m_0^{(1)}(x,z)=m_0^{(x)}(x)\,m_0^{(z)}(z)$)
and the fact that the underlying effective planar wall contribution
is a solution of the corresponding Euler-Lagrange equation Eq.~\eqref{eq:diffeq}, 
Eq.~\eqref{eq:m0-deq} turns into
\begin{equation} 
	\label{eq:m0-deq-2}
	-\frac{\frac{\partial^2}{\partial x^2}m^{(x)}_0(x)}{m^{(x)}_0(x)}\simeq
	\frac{\frac{\partial^2}{\partial z^2}m^{(z)}_0(z)}{m^{(z)}_0(z)} 
	-  3\left(m^{(0)}_0(z)\right)^2.
\end{equation}
Equation \eqref{eq:m0-deq-2} holds for $x$ and $z$ independently and therefore
both sides of Eq.~\eqref{eq:m0-deq-2} are equal to the same constant $k^2$ so that for the part of 
Eq.~\eqref{eq:m0-deq-2} depending on $x$ one has
\begin{equation} 
	\label{eq:mx-deq}
	\frac{\partial^2}{\partial x^2}m^{(x)}_0(x)=-k^2 m^{(x)}_0(x). 
\end{equation}
A solution of Eq.~\eqref{eq:mx-deq}, which is compatible with the periodic boundary conditions of the 
corrugated system [Fig.~\ref{fig:definitions}], is
\begin{equation} 
	m^{(x)}_0(x)=-M_x \cos(kx),
\end{equation}
where $k=\hat{k} \frac{2\pi}{\lproj}$ with $\hat{k}$ a natural number; $M_x$ is a positive amplitude.
We assume that far from the substrate only the main contribution due to the corrugation
is significant, i.e., $\hat{k}=1$ and $k={2\pi}/{\lproj}$.
%
For the $z$-dependent part of Eq.~\eqref{eq:m0-deq-2} one has
\begin{equation} 
	\label{eq:mz-deq}
	 \frac{\partial^2}{\partial z^2}m^{(z)}_0(z)=\left[3\left(m^{(0)}_0(z)\right)^2+k^2\right]m^{(z)}_0(z).
\end{equation}
Far from the substrate (i.e., $z\gg \lproj$) the underlying effective planar wall contribution
Eq.~\eqref{eq:m00-heff} fulfils
\begin{equation}
	 \left(m^{(0)}_0(z) \right)^2 \ll \frac{1}{3}\left(\frac{2\pi}{\lproj}\right)^2.
\end{equation}
Thus, together with $k=2\pi/\lproj$ from above, Eq.~\eqref{eq:mz-deq} can be written as
\begin{equation} 
	 \label{eq:mz-deq2}
	 \frac{\partial^2}{\partial z^2}m^{(z)}_0(z)\simeq\left(\frac{2\pi}{\lproj}\right)^2 m^{(z)}_0(z),
\end{equation}
with the physically relevant solution
\begin{equation}
	m^{(z)}_0(z)= M_z \exp\left\{-\frac{2\pi}{\lproj} z\right\}
\end{equation} 
and $M_z$ as a positive amplitude.
Accordingly, at criticality the contribution of the corrugation to the order parameter
decays \emph{exponentially} into the bulk, in contrast to the total order parameter, which
decays algebraically.
Upon inserting the effective planar wall solution, 
the final approximative expression for the order parameter reads
\begin{equation}
	\label{eq:approximation-zero-final}
	m_0(x,z)\simeq 
					\frac{c_+}{z-\heff}
					\,-\, M_{xz}\, \cos\left(\tfrac{2\pi}{\lproj} x \right)%
					 e^{-\frac{2\pi}{\lproj} z}, %
\end{equation}
where $M_{xz}=M_xM_z$.
$M_{xz}$ depends only on the corrugation height ($\propto 1/h$) and on the wedge angle 
$\gamma$ because for ${T=T_c}$ the product $m\times h$ depends only on $x/h$, $z/h$, and $\gamma$ 
[Eq.~\eqref{eq:scaling-zero}].
\par
From Eq.~\eqref{eq:approximation-zero-final} we calculate the undulation  $\Delta s(s)$ of the contour lines,
which has been defined in Subsec.~\ref{sec:undulation} as the difference of the maximum and the 
minimum value of $z/h$ of a given contour line of the order parameter.
The positions of the maximum and the minimum $z_\Max=h(s+\frac{1}{2}\Delta s)$ and 
$z_\Min=h(s-\frac{1}{2}\Delta s)$ of a contour line
correspond to the same $x$ values as the crest of an edge or the apex of a wedge
of the periodic array of wedges and ridges, i.e., $x=\lproj/2$ or $x=0$.
Thus, for any contour line far from the substrate one finds
\begin{widetext}
\begin{equation}
	\label{eq:contour-1}
	m_0(x=\lproj/2,z=z_\Max)=
		c_+ \,\left[h\left(s+\frac{1}{2}\Delta s - \hefftilde\right)\right]^{-1}
		+ M_{xz}\, e^{-\frac{2\pi h}{\lproj}(s+\frac{1}{2}\Delta s)}
\end{equation}
and
\begin{equation}
	\label{eq:contour-2}
	m_0(x=0,z=z_\Min)=
		c_+ \,\left[h\left(s-\frac{1}{2}\Delta s - \hefftilde\right)\right]^{-1}
		- M_{xz}\, e^{-\frac{2\pi h}{\lproj}(s-\frac{1}{2}\Delta s)}.
\end{equation}
\end{widetext}
Along a contour line the order parameter value is constant which allows one to equate the right-hand sides
of Eqs.~\eqref{eq:contour-1} and \eqref{eq:contour-2}:
\begin{equation}
	\label{eq:spatvar-zeroDs1}
	\frac{2\,M_{xz}\,h}{c_+\, e^{\frac{2\pi h}{\lproj}s}}\, 
	\cosh\left(\frac{\pi h}{\lproj}\Delta s\right)
	=  \frac	{\Delta s}
		{\left(s-\hefftilde\right)^2-\left(\frac{1}{2}\Delta s\right)^2}.
\end{equation}
Far from the wall $\Delta s(s)\ll 1$, and quadratic and higher order terms of $\Delta s(s)$ can be neglected.
With the additional assumption that $h$ is not much larger than $\lproj$,
one can approximate $\cosh\left(\frac{\pi h}{\lproj}\Delta s\right)\simeq 1$ and 
Eq.~\eqref{eq:spatvar-zeroDs1} becomes
\begin{equation}
	\label{eq:spatvar-zero-final}
	 \Delta s(s) =2\,\frac{M_{xz}\,h}{c_+}{\left(s-\hefftilde\right)^2} 
	 e^{-\frac{2\pi h}{\lproj}s}.
\end{equation}
Thus, the undulation of the contour lines of the order parameter decays exponentially into the bulk with the
decay constant
\begin{equation}
\sigma_0(\gamma)\equiv\frac{2\pi\,h}{\lproj}=\pi\cot(\gamma/2).
\end{equation}

\subsection{Distant behavior of the order parameter profile for ${T>T_c}$ \label{sec:appendix-pos}}
%
The order parameter profile for $T > T_c$ is characterized by the scaling function 
$P_+(x_+,z_+,h_+,\gamma)$ [Eq.~\eqref{eq:scaling}] which we abbreviate as $P_+(x_+,z_+)$.
The procedure follows the one of Subappendix~\ref{sec:appendix-zero} for ${T=T_c}$.
The MFT Euler-Lagrange equation for the scaling function reads
\begin{equation}
	\label{eq:P+deq}
	\left(\frac{\partial^2}{\partial z_+^2}+\frac{\partial^2}{\partial x_+^2}\right)  P_+(x_+,z_+)
	= P_+(x_+,z_+) +  P_+^3(x_+,z_+).
\end{equation}
According to Eq.~\eqref{eq:superposition} one separates $P_+$ into an effective planar wall 
contribution and a corrugation contribution:
\begin{equation} 
	\label{eq:spatvar-pos-ansatz}
	P_+(x_+,z_+)\simeq P_+^{(0)}(z_+)+P_+^{(1)}(x_+,z_+),
\end{equation}
where $P_+^{(0)}(z_+)$ is the effective planar wall MFT solution (see Eqs.~(\ref{eq:def-hefftilde}--
\ref{eq:definition-heff}) and Fig.~\ref{fig:cut}):
\begin{equation}
	\label{eq:P+eff}
	P_+^{(0)}(z_+)=\sqrt{2}\big/\sinh(z_+-h_+\,\hefftilde),
\end{equation}
which corresponds to the bulk value $P_+^{(0)}(z_+\to\infty)=0$.
Far from the substrate $P_+^{(1)}\ll P_+^{(0)}$, so that one can neglect terms of quadratic
or higher order in $P_+^{(1)}$.
Using the product ansatz according to Eq.~\eqref{eq:product}
(i.e., $P_+^{(1)}(x_+,z_+)= P_+^{(x)}(x_+)\, P_+^{(z)}(z_+)$),
upon inserting Eq.~\eqref{eq:spatvar-pos-ansatz} into Eq.~\eqref{eq:P+deq}
one finds the following differential equations:
\begin{equation} 
	\label{eq:Pxdeq}
	-{\frac{\partial^2}{\partial x_+^2}P_+^{(x)}(x_+)} =  k^2 {P_+^{(x)}(x_+)}
\end{equation}
and
\begin{equation}
	\label{eq:Pzdeq}
	\frac{\frac{\partial^2}{\partial z_+^2}P_+^{(z)}(z_+)}{P_+^{(z)}(z_+)} - 1 - 3(P_+^{(0)}(z_+) )^2
	= k^2.
\end{equation}
Similar to Subappendix~\ref{sec:appendix-zero} the solution of Eq.~\eqref{eq:Pxdeq} is
\begin{equation} 
	P_+^{(x)}(x_+)=-P_x^+\, \cos(k\,x_+)  
\end{equation}
where $P_x^+>0$ and $k=2\hat{k}\pi \xi_+/\lproj$ with a natural number $\hat{k}$.
By keeping only the main contribution $\hat{k}=1$, Eq.~\eqref{eq:Pzdeq} becomes
\begin{equation} 
	 \label{eq:spatvar-pos-deqz}
	 \frac{\partial^2}{\partial z_+^2}P_+^{(z)}(z_+)=\left(  1 +\left(\frac{2\pi\xi_+}{\lproj}\right)^2\right)P_+^{(z)}(z_+) 
\end{equation}
where $P_+^{(0)}(z_+)$ is replaced by its bulk value zero because $P_+^{(0)}(z_+)\ll1$
for large distances $z\gg h,\xi_+$.
Equation \eqref{eq:spatvar-pos-deqz} is solved by
\begin{equation}
	P_+^{(z)}(z_+)= P_z^+ e^{-\sqrt{ 1 +\left(\frac{2\pi\xi_+}{\lproj}\right)^2}\, \, z_+},
\end{equation}
where $P_z^+$ is a positive constant.
%
Finally, for $z_+\gg1$ the approximate solution for $P_+$ reads
\begin{equation}
	P_+(x_+,z_+)  \simeq P_+^{(0)}(z)  
	-P_x^+P_z^+\, e^{-\sqrt{1+\left(\frac{2\pi\xi_+}{\lproj}\right)^2}\, \, z_+}\, \cos\left(\tfrac{2\pi\xi_+}{\lproj} x_+ \right),
\end{equation}
with $P_+^{(0)}$ given by Eq.~\eqref{eq:P+eff}. For $z_+\gg1$ one has
\begin{equation}
	P_+^{(0)}(z_+\gg1)\propto \exp(-z_+).
\end{equation}
\par
Similar to Subappendix~\ref{sec:appendix-zero} one can directly identify the minimum and the maximum of the 
contour lines with
$(x_+=0\;,\;z_{+\,min}=h_+(s-\frac{1}{2}\Delta s))$ and
$(x_+=\frac{1}{2\xi_+}\lproj\;,\;z_{+\,max}=h_+(s+\frac{1}{2}\Delta s))$.
Using the abbreviation 
\begin{equation}
Q = h_+\,\sqrt{1 +\left(\frac{2\pi\xi_+}{\lproj}\right)^2},
\end{equation}
the comparison of the order parameter at these two positions yields
\begin{equation}
	\label{eq:spatvar-pos-Ds1}
	 Z e^{-Qs}\,\cosh\left(\frac{Q}{2}\Delta s\right)=e^{-h_+s}\,\sinh\left(\frac{h_+}{2}\,\Delta s\right),
\end{equation}
where $Z$ is a prefactor, which is independent of $s$ and includes all possible other 
prefactors.
Far from the substrate $\Delta s\ll1$, and one can neglect terms of quadratic or higher order of it.
With the additional assumptions that $h_+$ is not much larger than unity and $\lproj$ is not much smaller
than $\xi_+$ (i.e., $Q$ is not much larger than $1$) Eq.~\eqref{eq:spatvar-pos-Ds1} simplifies and one finds
\begin{equation}
	\Delta s (s) \propto  \exp\left\{-\left(Q-h_+\right)s\right\},
\end{equation}
where the proportionality constant depends on $h_+$ and $\gamma$.
Thus, for ${T>T_c}$ the undulation decays exponentially into the bulk with the decay constant
\begin{equation}
\sigma_+\left({h}_+,\gamma\right) = h_+
	\left[ \sqrt{ 1 + \left( \frac{1}{h_+} \pi \cot \left(\frac{\gamma}{2}\right) \right)^2} - 1 \right]. 
\end{equation}
\subsection{Distant behavior of the order parameter for ${T<T_c}$ \label{sec:appendix-neg}}
%
For ${T<T_c}$ one has to consider the scaling function $P_-(x_-,z_-,h_-,\gamma)$
[Eq.~\eqref{eq:scaling}] abbreviated as $P_-(x_-,z_-)$.
It obeys the differential equation [Eq.~\eqref{eq:diffeq}]
\begin{equation}
\label{eq:neg_deq}
 \left(\frac{\partial^2}{\partial z_-^2}+\frac{\partial^2}{\partial x_-^2}\right)  P_-(x_-,z_-)
= -\frac{P_-(x_-,z_-)}{2} + \frac{P_-^3(x_-,z_-)}{2}.
\end{equation}
Following an analogous procedure as in Subappendices~\ref{sec:appendix-zero} and \ref{sec:appendix-pos}
and using the ansatz $P_-=P_-^{(0)}+P_-^{(x)}\, P_-^{(z)}$ [Eqs.~\eqref{eq:superposition} and 
\eqref{eq:product}] one finds
\begin{equation} 
	\label{eq:spatvar-neg-deq}
	-\frac{\frac{\partial^2}{\partial x_-^2}P_-^{(x)}(x_-)}{P_-^{(x)}(x_-)}
	= 
	k^2 
	= \frac{\frac{\partial^2}{\partial z_-^2}P_-^{(z)}(z_-)}{P_-^{(z)}(z_-)} 
	  +\frac{1}{2}  -\frac{3}{2}(P_-^{(0)}(z_-) )^2.
\end{equation}
The part of Eq.~\eqref{eq:spatvar-neg-deq} depending on $x_-$ is solved by
\begin{equation} 
	P_-^{(x)}(x_-)=-P_x^-\, \cos(kx_-), 
\end{equation}
where $P_x^->0$ and $k= \frac{2\pi\xi_-}{\lproj}$.
Accordingly, the part of Eq.~\eqref{eq:spatvar-neg-deq} depending on $z_-$ reads
\begin{equation} 
	 \frac{\frac{\partial^2}{\partial z_-^2}P_-^{(z)}(z_-)}{P_-^{(z)}(z_-)}
	 =
	 \frac{3}{2}(P_-^{(0)}(z_-))^2- \frac{1}{2} +\left(\frac{2\pi\xi_-}{\lproj}\right)^2. 
\end{equation}
Far from the wall one can replace $P_-^{(0)}(z_-\gg1)$ by $1$:
\begin{equation}
	\label{eq:repkace}
	\frac{3}{2}(P_-^{(0)}(z_-))^2-\frac{1}{2} +\left(\frac{2\pi\xi_-}{\lproj}\right)^2\simeq
	1 +\left(\frac{2\pi\xi_-}{\lproj}\right)^2.
\end{equation}
With this replacement Eq.~\eqref{eq:spatvar-neg-deq} is the same as 
Eq.~\eqref{eq:spatvar-pos-deqz} for ${T>T_c}$, with $\xi_-$ replacing $\xi_+$.
Thus, with Eq.~\eqref{eq:scaling-decay-2} the undulation for ${T<T_c}$ can be approximated by
\begin{equation}
\Delta s(s)\propto e^{-\sigma_-\left(h_-,\gamma\right)s},
\end{equation}
where
\begin{equation}
\sigma_-\left(h_-,\gamma\right)=h_-\left[
	\sqrt{1+\left(\frac{1}{h_-}\pi\cot\left(\frac{\gamma}{2}\right) \right)^2} -1 \right]. 
\end{equation}
%
\section{Distant behavior of the lateral critical Casimir force for ${T=T_c}$\label{sec:appendix-2}}
%
In this appendix we discuss the distant wall behavior of the lateral critical Casimir force between 
two identically structured substrates (see Subsec.~\ref{sec:lateral}) by using a superposition approximation
based on Eqs.~\eqref{eq:superposition} and \eqref{eq:product} and Appendix~\ref{sec:appendix-1}.
\par
In Subappendix~\ref{sec:appendix-zero} the corrugation contribution $m_0^{(1)}$ to the order parameter 
$m_0\simeq m_0^{(0)}+m_0^{(1)}$ for ${T=T_c}$ close to a single corrugated wall 
[Fig.~\ref{fig:definitions}] has been approximated 
for large distances from the wall as [Eq.~\eqref{eq:approximation-zero-final}]
\begin{equation}
	\label{eq:corrugation-contribution}
	m_0^{(1)}(x,z)=  -\, M_{xz}\, 
					\cos\left(\tfrac{2\pi}{\lproj} x \right)%
					 e^{-\frac{2\pi}{\lproj} z}, %
\end{equation}
where $M_{xz}$ is a positive number which only depends on $h$ as $\propto h^{-1}$ and $\gamma$.
\par
For the $\Ss$ geometry [Fig.~\ref{fig:defstrucstruc}] one has to deal with the corrugations
of both walls located at $z=0$ and $z=L$ as well as with the lateral shift $D$ along the $x$ direction.
The corrugation contribution of the lower and upper wall in Fig.~\ref{fig:defstrucstruc} to the order
parameter is $m_0^{(1)}(x,z)$ [Eq.~\eqref{eq:corrugation-contribution}] and $m_0^{(1)}(x-D,L-z)$, 
respectively.
The underlying planar wall contribution for the $\Ss$ geometry is the order parameter profile
$m_0^{(0)}=m_0^{+,+}(z,\Leff)$ between two planar walls at distance $\Leff$ in the
strong adsorption limit for identical chemical boundary conditions \cite{KRECH97};
$m_0^{+,+}(z,\Leff)$ does not depend on the lateral position variable $x$.
In order to determine the distant wall behavior it is convenient to approximate the total order 
parameter at ${T=T_c}$ in the $\Ss$ geometry as
\begin{widetext}
\begin{multline}
	m_0(x,z,L\gg h,D,h,\gamma) \simeq m_0^{+,+}(z,\Leff) \,+\, m_0^{(1)}(x,z) \,+\, m_0^{(1)}(x-D,L-z)\\
		= m_0^{+,+}(z,\Leff) \,-\, M_{xz}  \left\{
					\cos\left(\tfrac{2\pi}{\lproj} x \right)%
					 e^{-\frac{2\pi}{\lproj} z}\,+\, %
					\cos\left(\tfrac{2\pi}{\lproj} (x-D) \right)%
					 e^{-\frac{2\pi}{\lproj} (L-z)} %
				\right\}.
\end{multline}
The lateral critical Casimir force is given by the integral over the stress tensor component $T_{xz}$
(see Eqs.~\eqref{eq:stresstensor} and \eqref{eq:stresstensor-2}):
\begin{equation}
	\label{eq:forceintegral}
	f_\parallel^\Ss(D,L,h,\gamma,t=0)=
			\frac{3!}{u\;\lproj}
			\int\limits_0^{\lproj}
			dx\;
				  \left.\frac{\partial m_0(x,z,L,D,h,\gamma)}{\partial z}\right|_{z=z_0}
				\,\left.\frac{\partial m_0(x,z,L,D,h,\gamma)}{\partial x}\right|_{z=z_0},
\end{equation}
\end{widetext}
where $z_0$ is an arbitrary but fixed position in between the two substrates.
For simplicity we choose $z_0=L/2$, i.e., the center position in between the two substrates, where
the planar wall contribution $m_0^{+,+}$ to the order parameter takes its minimal value, i.e.,
\begin{equation}
	\left(\frac{\partial m_0^{+,+}(z,\Leff)}{\partial z}\right)_{z=\frac{L}{2}}=
	\frac{\partial m_0^{+,+}(z,\Leff)}{\partial x}=0.
\end{equation}
With this the integrand in Eq.~\eqref{eq:forceintegral} reduces to a product of the
derivatives of the sum of the corrugation contributions.
Performing the integration in Eq.~\eqref{eq:forceintegral} leads to Eq.~\eqref{eq:caslat-final} with the 
dimensionless amplitude 
\begin{equation}
	\label{eq:K-def}
	K  = 4\pi^2\,\frac{3!}{u}\,|\Delta_{+,+}|^{-1}\,\frac{M_{xz}^2\,h^d}{\lproj^2},
\end{equation}
where $d=4$ within MFT. 
Note that $\lproj=2h\tan(\gamma/2)$ and $M_{xz}$ is inversely proportional to $h$; 
accordingly, $K$ is positive and depends only on $\gamma$.
\section{Comparison of the lateral critical Casimir force with lateral van der Waals forces\label{sec:vdw}}
\subsection{Lateral van der Waals forces for geometrically structured confinements within the
approximation of summing pair potentials\label{sec:pws}}
Van der Waals or dispersion forces provide a nonsingular background contribution which
adds to the critical Casimir forces.
For example, the particles constituting the confining substrates in the $\Ss$ 
geometry [Fig.~\ref{fig:defstrucstruc}]
do not only interact with the fluid, but also among each other, such that
their inhomogeneous distribution gives rise to a lateral force acting on the substrates even in
the absence of the fluid in between them.
The strength of these background forces can be determined by analyzing the confining system
\emph{without} a fluid.
\par
In order to estimate the contribution of the background interaction between the substrate molecules
to the total force, we assume their corresponding pair potential to be given by the well known 
non-retarded van der Waals potential,
\begin{equation}
	u(\vec{r},\vec{r'}) =
	-\frac{\mathcal{\hat{A}}}{\pi^2\rho\rho'}\,\frac{1}{|\vec{r}-\vec{r'}|^6},
\end{equation}
where $\vec{r}=(x,y,z)$ and $\vec{r'}=(x',y',z')$ are the three-dimensional position vectors 
of the two interacting molecules,
the amplitude $\mathcal{\hat{A}}$ characterizing the strength of the interaction between the molecules
is chosen such that it turns into the Hamaker constant, and
$\rho$ and $\rho'$ are the number densities of the particles in the two interacting bodies
(here $\rho'=\rho$ because we consider identical substrate materials).
\par
We use the pairwise summation approximation (PWS) in order to calculate the total
van der Waals interaction potential between two macroscopic bodies,
\begin{equation}
	\label{eq:vdwint}
	U^{vdW}\simeq-\int_B\int_{B'}\,d^3\vec{r}\,d^3\vec{r'}\,\rho\rho'u(\vec{r},\vec{r'}), 
\end{equation}
where $B$ and $B'$ denote the volumes of the two interacting substrates.
Note that for the geometries under consideration, the typical sizes of which are large on 
molecular scales, assuming a non-retarded van der Waals potential as well as the PWS amount to a severe 
approximation.
But here using them is appropriate because we want to obtain a simple comparison between the strengths of 
the lateral van der Waals force and the lateral critical Casimir force for the $\Ss$ geometry 
[Fig.~\ref{fig:defstrucstruc}].
\par
The van der Waals interaction potential per area $A=\lproj\times l_y$ within the $xy$ plane 
corresponding to the lateral van der Waals force for 
the $\Ss$ geometry is given by
\begin{widetext}
\begin{equation}
	\label{eq:vdwint-2}
	\frac{U^{vdW}_\parallel}{A}=
		-\frac{\mathcal{\hat{A}}}{\pi^2 A}
		\int\limits_{-\lproj/2}^{\lproj/2}\!\!\!\!dx
		\int\limits_0^{l_y}dy\!\!\!
		\int\limits_0^{z_\Max(x)}\!\!\!dz
		\int\limits_{-\infty}^{\infty}dx'
		\int\limits_{-\infty}^{\infty}dy'\!\!\!
		\int\limits_{{z}_\Min(x')}^{L}\!\!\!dz'
		\,\frac{1}{\left((x-x')^2+(y-y')^2+(z-z')^2\right)^3},
\end{equation}
where $z_\Max(x)$ is the $z$ value as a function of $x$ of the lower substrate surface in 
Fig.~\ref{fig:defstrucstruc}, 
\begin{equation}
	\label{eq:zmax}
	z_\Max(x)=
	\begin{cases}
		\frac{2\,h}{\lproj}|x|\,,\qquad&-\frac{\lproj}{2}\le x\le\frac{\lproj}{2},\\
		z_\Max(x\mp \lproj),&x\gtrless\pm \frac{\lproj}{2},
	\end{cases}
\end{equation}
and ${z}_\Min(x)$ is the corresponding $z$ value of the upper substrate surface.
For identical but shifted corrugations one has ${z}_\Min(x)=L-z_\Max(x-D)$, 
where $D$ is the lateral shift between the two substrates along the $x$ direction 
[Fig.~\ref{fig:defstrucstruc}].
In Eq.~\eqref{eq:vdwint-2} only the volumina of the corrugations on top of the two planar substrates have 
been considered.
$U^{vdW}-U_\parallel^{vdW}$ is the interaction potential between two planar semi-infinite substrates at 
distance $L$; it depends only on $z$ and therefore does not give rise to a lateral force.
With $\int_{-\infty}^{\infty}dy(a^2+y^2)^{-3}=3\pi/(8a^5)$, the integral in Eq.~\eqref{eq:vdwint-2} 
reduces to
\begin{equation}
	\label{eq:vdwint-3}
	\frac{U^{vdW}_\parallel}{A}=
	-\frac{3\,\mathcal{\hat{A}}}{8\pi\lproj}
		\int\limits_{-\lproj/2}^{\lproj/2}\!\!\!\!dx\!\!\!\int\limits_0^{z_\Max(x)}\!\!\!dz
		\int\limits_{-\infty}^{\infty}dx'\!\!\!\int\limits_{L-z_\Max(x'-D)}^{L}\!\!\!dz'
		\,\frac{1}{\left((x-x')^2+(z-z')^2\right)^{5/2}}.
\end{equation}
The lateral van der Waals force $f^{vdW}_\parallel$ per area corresponds to the derivative of 
$U^{vdW}_\parallel/A$ with respect to the lateral shift $D$,
\begin{equation}
	\label{eq:vdwint-4}
	f_\parallel^{vdW}
	=
	-\frac{\partial}{\partial D}\,\frac{U_\parallel^{vdW}}{A}
	=
	-	\frac{3\,\mathcal{\hat{A}}}{8\pi\lproj}
		\int\limits_{-\lproj/2}^{\lproj/2}\!\!\!\!dx
		\!\!\!\int\limits_0^{z_\Max(x)}\!\!\!dz
		\int\limits_{-\infty}^{\infty}dx'
		\,\frac{{z'}_\Max(x'-D)}{\left((x-x')^2+(z+z_\Max(x'-D)-L)^2\right)^{5/2}},
\end{equation}
where ${z'}_\Max(x)=\frac{d}{dx}z_\Max(x)$.
Rescaling the lengths in the integrals in Eq.~\eqref{eq:vdwint-4} with the projected width $\lproj$,
\begin{equation}
	\widetilde{x}=x/\lproj,\qquad\widetilde{x}\,'=x'/\lproj,\qquad\widetilde{z}=z/\lproj,
	\qquad \delta=D/\lproj, \qquad \widetilde{L}=L/\lproj,
\end{equation}
and defining the dimensionless function $\widetilde{z}_\Max$ 
(corresponding to Eq.~\eqref{eq:zmax} and with $h=\lproj\cot(\gamma/2)/2$ [Fig.~\ref{fig:definitions}]) as
\begin{equation}
	\label{eq:zmax-1}
	\widetilde{z}_\Max(\widetilde{x},\gamma)=
		\begin{cases}
		\cot(\gamma/2)|\widetilde{x}\,|\,,\qquad & -\frac{1}{2}\le \widetilde{x} \le\frac{1}{2},\\
		\widetilde{z}_\Max(\widetilde{x}\mp 1), & \widetilde{x}\gtrless\pm \frac{1}{2},
		\end{cases}
\end{equation}
the lateral van der Waals force [Eq.~\eqref{eq:vdwint-4}] reads
\begin{equation}
	\label{eq:vdwint-5}
	f^{vdW}_\parallel\simeq
	-
	\frac{3\,\mathcal{\hat{A}}}{8\pi}\,\left(\frac{1}{\lproj}\right)^3
	\widetilde{f}_\parallel^{vdW}(\delta,\widetilde{L},\gamma),
\end{equation}
where $\widetilde{f}_\parallel^{vdW}$ plays the role of a dimensionless scaling function of
the lateral van der Waals force for the $\Ss$ geometry:
\begin{equation}
	\label{eq:vdwscaling}
	\widetilde{f}_\parallel^{vdW}(\delta,\widetilde{L},\gamma)=
	\int_{-1/2}^{1/2}\limits d\widetilde{x}\!\!\!
	\int\limits_0^{\widetilde{z}_\Max(\widetilde{x},\gamma)}\!\!\!d\,\widetilde{z}\,
	\int\limits_{-\infty}^{\infty}d\widetilde{x}\,'
	\,
	\frac {{\widetilde{z}\,'\!\!}_\Max(\widetilde{x}\,'-\delta,\gamma)}
		  {\left((\widetilde{x}-\widetilde{x}\,')^2+(\widetilde{z}+
		  \widetilde{z}_\Max(\widetilde{x}\,'-\delta,\gamma)-\widetilde{L})^2\right)^{5/2}},
\end{equation}
with ${\widetilde{z}\,'\!\!}_\Max(\widetilde{x},\gamma)=\frac{d}{d\widetilde{x}}z_\Max(\widetilde{x},\gamma)$.
\end{widetext}
\subsection{Numerical comparison \label{sec:num}}
%
In order to gauge the strength of the lateral critical Casimir force [Subsec.~\ref{sec:lateral}], we 
compare its amplitude with the one of the lateral van der Waals background force obtained within the 
PWS approximation [Subappendix~\ref{sec:pws}] for a specifically chosen geometry.
\par
For the amplitude of the lateral critical Casimir force for the $\Ss$ geometry we consider at ${T=T_c}$
the specific configuration $\gamma=90\degree$ and $L/h=4$ (i.e., $\Ltip/h=2$) [Fig.~\ref{fig:defstrucstruc}].
For this choice the corresponding lateral critical Casimir force scaling function amplitude has 
approximately the value (see the solid curve with square symbols in Fig.~\ref{fig:lateral-amp-scaled}(b))
\begin{equation}
	\widetilde{\Delta}_{\parallel,\Max}^\Ss(L/h=4,\gamma=90\degree)\times[\Ltip/L]^d \simeq 
		4.1\times10^{-3}.
\end{equation}
The corresponding maximum of the lateral critical Casimir force is (see Eq.~\eqref{eq:crit-force-ss-1})
\begin{equation}
	\label{eq:lat-force-1}
	f_{\parallel,\Max}^\Ss=|\Delta_{+,+}|\frac{d-1}{\Ltip^{d}}\,\left\{
	\widetilde{\Delta}^\Ss_{\parallel,\Max}(L/h,\gamma)\,\frac{\Ltip^d}{L^d}\right\}.
\end{equation}
The lateral critical Casimir force $f_\parallel^\Ss$ is given in units of the characteristic thermal 
energy $k_B T_c$ and the $d-1$ dimensional cross-sectional area $A$ within the $x\vec{y}$ plane.
Accordingly, the maximum of the lateral critical Casimir force per area 
$f^{cCas}_{\parallel,\Max}(L/h,\gamma)$ at \emph{c}riticality for the configuration chosen above is given by
\begin{equation}
	\label{eq:lat-force-2}
	f_{\parallel,\Max}^{cCas}(L/h=4,\gamma=90\degree)= k_B T_c\, f_{\parallel,\Max}^\Ss.
\end{equation}
Although the scaling function $\widetilde{\Delta}_\parallel^\Ss$ was determined within mean field theory,
its form gives a reasonable approximation for dimensions $d<d_{uc}=4$, too, if one replaces all
quantities depending on the dimension of the system by their corresponding value for the actual dimension $d$.
In order to obtain an estimate of the strength of the lateral critical Casimir force in $d=3$
we use $|\Delta_{+,+}|=0.44$ \cite{Vasilyev:2007}.
As typical length and temperature scales we use $L=200$\,nm (i.e., $h=50$\,nm, $\Ltip=100$\,nm, and
$\lproj=100$\,nm for the specific geometry under consideration) and we take $T_c=300$\,K.
With these values, the maximum of the lateral critical Casimir force [Eq.~\eqref{eq:lat-force-2}] is of the 
order of $f_{\parallel,\Max}^{cCas} \simeq 15$\,fN$/(\mu$m$)^2$.
\par
In order to estimate the strength of the contribution of the background forces to the total force
we use the PWS approximation result for the lateral dispersion force for the $\Ss$ geometry 
[Subappendix~\ref{sec:pws}].
For the lateral van der Waals force scaling function for the geometry defined above ($L/h=4$, 
$\gamma=90\degree$, i.e., $\widetilde{L}=L/\lproj=2$) one has
[Eq.~\eqref{eq:vdwscaling}]
\begin{equation}
	\widetilde{f}_\parallel^{vdW}(\delta=0.25,\widetilde{L}=2,\gamma=90\degree) \simeq 1.3\times10^{-4}.
\end{equation}
Typical values of the Hamaker constant $\mathcal{\hat{A}}$ are of the order of $10 k_B T$.
Using $\lproj=100$\,nm as above, we find for the maximum of the lateral van der Waals force 
[Eq.~\eqref{eq:vdwint-5}] for the $\Ss$ geometry a typical value of 
$|f^{vdW}_{\parallel,\Max}| \simeq 0.64$\,fN$/(\mu$m$)^2$.
This means that the background van der Waals interaction calculated within the PWS approximation, which 
represents the background interaction, is of the order of $1\%$ to $10\%$ of the lateral critical Casimir 
force.
Thus, near criticality the lateral critical Casimir forces are expected to be much stronger than the 
background forces.
%

\end{document}